\documentclass{article}

 \usepackage[preprint]{neurips_2025}

\usepackage[utf8]{inputenc} % allow utf-8 input
\usepackage[T1]{fontenc}    % use 8-bit T1 fonts
\usepackage{hyperref}       % hyperlinks
\usepackage{url}            % simple URL typesetting
\usepackage{booktabs}       % professional-quality tables
\usepackage{amsfonts}       % blackboard math symbols
\usepackage{nicefrac}       % compact symbols for 1/2, etc.
\usepackage{microtype}      % microtypography
\usepackage{xcolor}         % colors
\usepackage{glossaries}
\usepackage[capitalize,noabbrev]{cleveref}
\usepackage{tikzducks}
\usepackage{algorithm,algpseudocode}
\usepackage{comment}
\usepackage{todonotes}
\usepackage{natbib}
\usepackage{subcaption}

\glsdisablehyper
\newacronym{AFB}{AFB}{Air Force Base}
\newacronym{AI}{AI}{artificial intelligence}
\newacronym{APL}{APL}{%
  the Johns Hopkins University Applied Physics Laboratory%
}

\newacronym{BO}{BO}{Bayesian optimization}

\newacronym{CMDP}{CMDP}{Constrained Markov Decision Process}
\newacronym{COAWST}{COAWST}{Coupled Ocean–Atmospheric–Wave–Sediment Transport}

\newacronym{DID}{DID}{defense in depth}
\newacronym{DL}{DL}{deep learning}
\newacronym{DRL}{DRL}{deep reinforcement learning}

\newacronym{EWE}{EWE}{extreme weather event}

\newacronym{GHSL}{GHSL}{Global Human Settlement Layer}
\newacronym{GP}{GP}{Gaussian process}

\newacronym{JHU}{JHU}{%
  Johns Hopkins University%
}

\newacronym{MAE}{MAE}{mean absolute error}
\newacronym{ML}{ML}{machine learning}
\newacronym{MSE}{MSE}{mean squared error}

\newacronym{NN}{NN}{neural network}
\newacronym{NO}{NO}{neural operator}
\newacronym{NTK}{NTK}{neural tangent kernel}
\newacronym{NTS}{NTS}{Neural Thompson Sampling}

\newacronym{PDE}{PDE}{partial differential equation}

\newacronym{RL}{RL}{reinforcement learning}
\newacronym{ROMS}{ROMS}{Regional Ocean Modeling System}

\newacronym{seq2seq}{seq2seq}{sequence-to-sequence}
\newacronym{SM}{SM}{surrogate model}
\newacronym{STO-BNTS}{STO-BNTS}{Sample-Then-Optimize Batch Neural Thompson Sampling}
\newacronym{SWAN}{SWAN}{Simulating WAves Nearshore}
\newacronym{SwinIR}{SwinIR}{shifted window image restoration}
\newacronym{SWL}{SWL}{still water level}

\newacronym{TCN}{TCN}{temporal convolutional network}

\newacronym{USGS}{USGS}{United States Geological Services}

\newacronym{WOT}{WOT}{wave over topping}
\newacronym{WRF}{WRF}{Weather Research and Forecast}

\title{Discovering strategies for coastal resilience with AI-based prediction and optimization}
\workshoptitle{AI4Science}

\author{%
  Jared Markowitz$^*$, Alexander New$^*$, Jennifer Sleeman$^*$\\
  $^*$ equal contribution\\
  Research and Exploratory Development Department\\
  Johns Hopkins Applied Physics Laboratory\\
  \texttt{\{jared.markowitz, alex.new, jennifer.sleeman\}@jhuapl.edu}\\
  \And Chace Ashcraft, Jay Brett, Gary Collins, Stella In, Nathaniel Winstead\\
  Johns Hopkins Applied Physics Laboratory\\
}

\begin{document}

\maketitle

\begin{abstract}
Tropical storms cause extensive property damage and loss of life, making them one of the most destructive types of natural hazards. The development of predictive models that identify interventions effective at mitigating storm impacts has considerable potential to reduce these adverse outcomes. In this study, we use an \gls{AI}-driven approach for optimizing intervention schemes that improve resilience to coastal flooding. We combine three different \gls{AI} models to optimize the selection of intervention types, sites, and scales in order to minimize the expected cost of flooding damage in a given region, including the cost of installing and maintaining the interventions. Our approach combines data-driven generation of storm surge fields, surrogate modeling of intervention impacts, and the solution of a continuum-armed bandit problem. We applied this methodology to optimize the selection of sea wall and oyster reef interventions near Tyndall \gls{AFB} in Florida, an area that was catastrophically impacted by Hurricane Michael. Our analysis predicts that intervention optimization could potentially be used to save billions of dollars in storm damage, far outpacing greedy or non-optimal solutions. 

%We have used an \gls{AI}-driven framework to predict optimized intervention schemes, given storm field properties and cost for improved resilience to coastal flooding. Our model solves the intervention-placement problem: deciding which interventions to deploy, where the interventions should go, and how large they should be. This will minimize expected flooding damage in the affected region, incorporating installation cost estimates for the interventions. We evaluated our methodology on a specific use case for optimizing the placement and height of a sea wall and oyster reefs near Tyndall \gls{AFB} in Florida, an area that was catastrophically impacted by Hurricane Michael. We find that interventions are predicted to affect projected flooding costs of billions of dollars per storm.

\end{abstract}

\glsresetall

\section{Introduction}\label{sec:intro}

% \alex{this is too long, also update it with final version from  abstract edits}

% \alex{if we're going to use the term ``defense in depth'' in the title, we should explain what that means here}
% \jay{I recommend not-- seems to be deprecated in cyber, EP4 committee was not staying strongly tied to it}

As climate change intensifies, coastal regions will become increasingly vulnerable to extreme weather~\cite{SanchezArchilla2016coasts,Lawrence2018coasts,Gargiulo2020coast,Storlazzi2025reefs}. Nascent computational and experimental results have demonstrated how targeted placement of interventions can attenuate flooding and reduce damage~\cite{Storlazzi2025reefs,suttongrier2015hybrid,currin2019shorelines,jordan2022bridging,Brett2024oysters,ceres2019coastaloptimization}. However, the relative impacts of different intervention types (e.g., nature-based solutions like oyster reefs and marshes, grey solutions like sea walls) is not fully understood. Furthermore, the limited representation of interventions in large weather datasets like ERA5~\cite{Hersbach2020era5} makes it difficult to robustly assess their effectiveness. 

We have developed an \gls{AI}-driven framework (\cref{fig:schematic}) to predict optimized intervention schemes for improved resilience to coastal flooding, given storm properties and the estimated costs of damage and interventions. We evaluated our methodology on a specific use case: optimizing the placement and height of a sea wall and the placement of oyster reefs near Tyndall \gls{AFB} in Florida, an area that was catastrophically impacted by Hurricane Michael~\cite{beven2019michael}. Damage from Hurricane Michael was responsible for 16 direct deaths and 43 indirect deaths in the United States, as well as approximately \$30 billion in damages~\cite{beven2019michael}. While our models suggest that optimized interventions may reduce flooding costs by billions of dollars per storm, we also find that na\"ive choices may direct water more toward populated areas and thus increase damage and costs.

Our approach uses physics-based modeling of the impact of oyster reefs on wave height and a sea wall on inland flooding
%~\cite{Warner2008coawst}
(\cref{sec:damage} and~\cref{sec:data}), \gls{AI} models for additional storm data generation (\cref{sec:generative_model}), \gls{AI}-based \glspl{SM}
%(SwinIR vision transformers~\cite{Liang2021swinir},~\cref{sec:img2img})
to rapidly predict how oyster reefs change wave height and direction, and identification of optimal interventions with black-box optimization
% via \gls{STO-BNTS}~\cite{sto_bnts} with prioritized sampling 
(\cref{sec:optimization}).  The framework allows one to decide which interventions to deploy, where they should go, and how large they should be. Selected interventions minimize the cost of expected flooding damage in the affected region, incorporating installation and maintenance costs. 

%\alex{@jennifer, do you want to add a one-sentence each summary of results to the following paragraph?}
%\jared{are we happy with presenting contributions as the following?}

Our key findings are that:
\begin{itemize}
\item The computational challenge of producing at-scale storm training data may be overcome by training a model (inspired by the U-Net~\cite{Ronneberg2015unet} architecture) to produce storm surge fields based on atmospheric and wave conditions.
%To overcome the computational challenges of producing at-scale training data using traditional numerical models for this problem setup, we explored reducing the complexity of a three-way coupled model (atmosphere, wave, and ocean) by using an \gls{AI} model in place of the surge from the ocean model.  This reduced the coupling to two models, providing significant speed-up in terms of data generation.  The surge generator model produces storm surge fields based on atmospheric and wave fields.

\item A surrogate model with transformer architecture~\cite{Liang2021swinir} may be used to accurately predict the effect of oyster reefs on wave heights (\cref{sec:surrogate_results}) while requiring only a fraction of the computational time of physics-based modeling.

\item The combination of offshore (oyster reef) and onshore (sea wall) interventions may be optimized~\cite{Dai2022sto_bnts} to potentially save billions of dollars per storm and to increase savings over a projected 50-year intervention lifetime by as many as tens of billions of dollars.
%Given these inputs, our optimization approach is able to identify interventions that would have saved approximately \$5.5B on flooding from Hurricane Michael, including the cost of their installation and maintenance.  On a broader set of  representative storms, our models suggest that (1) net financial savings from intervention installation should easily exceed \$100B over a 50-year intervention lifetime and (2) optimized intervention placement should add more than \$10B to lifetime savings.
\end{itemize}

While the specific numerical results presented are particular to the region surrounding Tyndall \gls{AFB}, we expect the approach to be applicable to many different sites and intervention types.

\subsection{Related work}\label{sec:literature}

% \alex{folks who have ready-made text should feel free to throw in here}

Recent weather and earth systems \glspl{SM}~\cite{Lam2023graphcast,Pathak2022fourcastnet,Nguyen2023climax,Bodnar2025aurora,Nagaraj2025surgemodeling} have demonstrated effective short and medium-term forecasts on global and regional scales. However, they are trained on coarse geospatial meshes and their predictions neither resolve the effects of small-scale interventions nor support the inclusion of interventions. 
%Additionally, having a sufficient representation of diverse off-shore storms globally is critical. 
Thus, being able to accurately and effectively model interventions requires the generation of a training dataset with different off-shore storms and interventions.

\Glspl{SM} that rely less on large datasets have also been applied to this problem. For example, ~\cite{Ions2024xbeachgpr} used \glspl{GP} to predict wave attenuation coefficients given the presence of rigid vegetation, while ~\cite{Mj2020bowaves} used \gls{BO} to identify wave conditions that optimize specified criteria. Similarly,~\cite{miura2021optimization} optimize for coastal protection based on outputs of physics-based calculations. These approaches and others have shown promise; however, they rely on defining a bespoke problem and then generating data specific to that problem. There is still a need for a general approach that can optimize intervention effects across geographies, storms, and intervention types.

\section{Technical approach}\label{sec:approach}

\begin{figure}
    \centering
    \fbox{\includegraphics[width=0.9\linewidth]{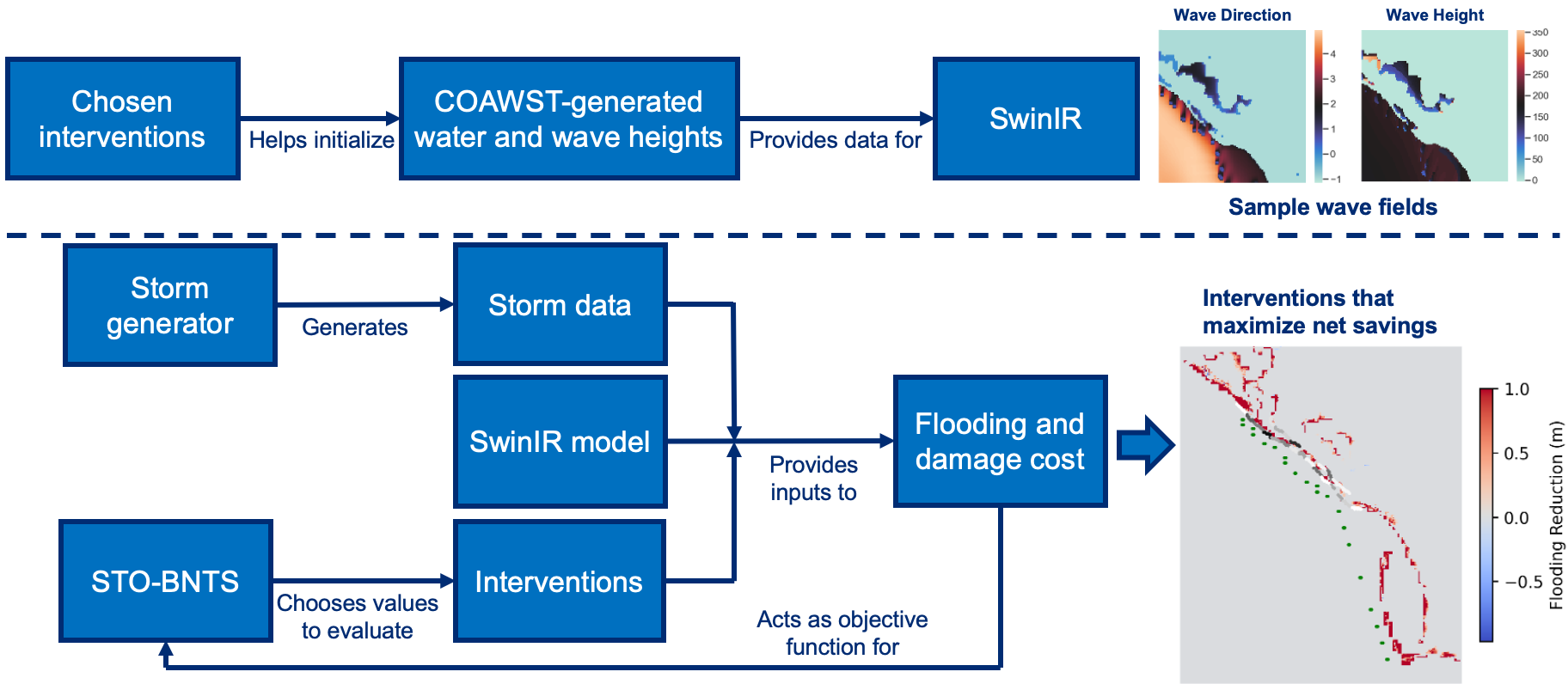}}
    % \begin{tikzpicture}
    %     \randuck
    % \end{tikzpicture}
    \caption{Our framework combines numerical and \gls{AI}-based storm data generation, shifted window image restoration transformers~\cite{Liang2021swinir} for predicting intervention effects, and Sample-Then-Optimize Batch Neural Thompson Sampling~\cite{Dai2022sto_bnts} 
    % with prioritized action initialization 
    for optimizing interventions.}
    \label{fig:schematic}
\end{figure}

\subsection{Physics-based storm data simulation}\label{sec:data}
% \alex{assigned to @pete/@jennifer}

\Gls{COAWST}~\cite{Warner2008coawst,warner2010development} is a \gls{USGS} modeling framework built to simulate the interactions between ocean, atmosphere, waves, and sediment transport processes. Using \Gls{COAWST}, we recreated Hurricane Michael. 
%at an 800 m resolution in the \gls{WRF} domain, with a horizontal grid resolution of approximately 300 m for the \gls{SWAN} and \gls{ROMS} domain. 
It took approximately one week using a high performance computing environment to generate 48 hours of storm data. %while allowing for spin-up of the modeled wind and wave fields
More details on our use of \gls{COAWST} are in~\cref{sec:coawst_extras}.

\begin{figure}
    \centering
    \includegraphics[width=\linewidth]{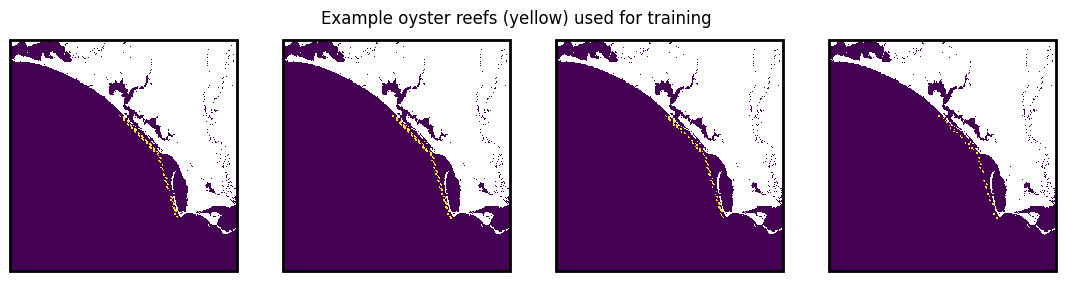}
    \caption{Four of the oyster reef configurations (yellow) for which we generated wave fields using \gls{COAWST} to use as training and validation data.
    % \alex{need to rotate these so they're correct}
    }
    \label{fig:reefs}
\end{figure}

For our Michael data, we additionally simulated wind and wave fields for different oyster reef configurations. Specifically, we adopted the approach of~\cite{Brett2024oysters} and used the vegetation dynamics module included with \Gls{COAWST} to treat oysters as a region of stiff plants located in a user-defined area. \Cref{fig:reefs} shows some of the oyster reef configurations for which we generated wave fields.

Unfortunately, storm data generation using traditional numerical models does not scale to the data needs of modern \gls{AI}. %The challenge in generating storm data for the purpose of training deep learning models is the limited volume of data that could be generated using traditional numerical models.  Given a three-way coupled system such as , this is particularly problematic.  
The \gls{COAWST} three-way coupled system tends to be prone to numerical instability and is difficult to configure in a high-performance computing environment.  For a single storm (using a fully coupled model), it could take a week of compute to simulate a two-day storm.  
%It is simply not an option for generating storm data at scale to use in training deep learning models.

For this reason, we explored the use of \gls{AI} models to decrease the needed \gls{COAWST} computation time. By reducing the coupling to a two-way coupling of atmospheric and wave models, we were able to reduce the storm simulation time by a factor of five.  The model generates the storm surge instead of using the coupled ocean model with the atmospheric and wave models.

\subsection{Quickly predicting high-resolution storm surge data from low-resolution data}\label{sec:generative_model}

\begin{figure}
    \centering
    \includegraphics[width=1.0\linewidth]{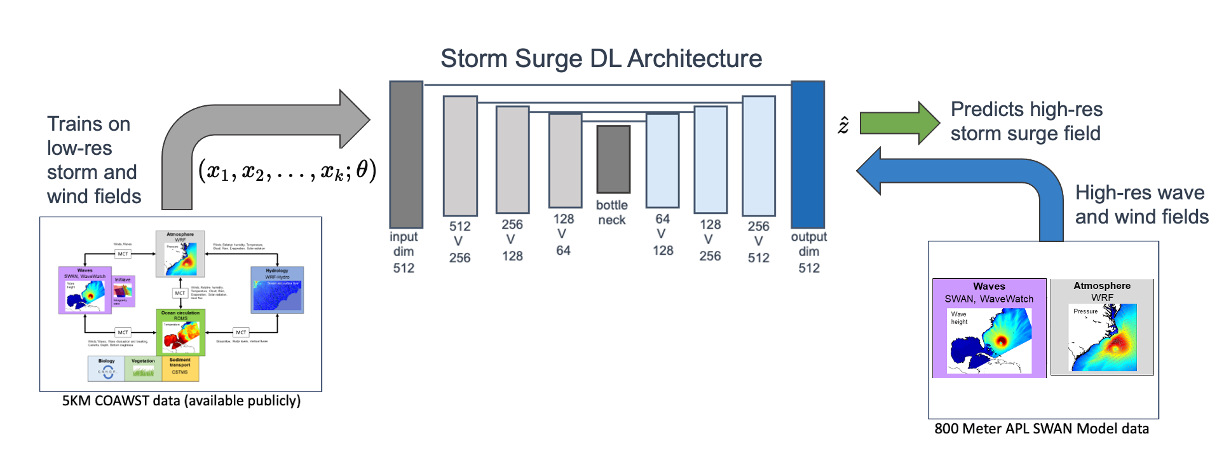}
    \caption{The Storm Surge Encoder Decoder Skip Connection model (\cref{sec:generative_model}). It is trained on 5 km COAWST data and used to predict storm surge at higher resolutions.}
    \label{fig:surge_model}
\end{figure}

% \alex{assigned to @jennifer}
To overcome the computational bottleneck of running the fully coupled ~\Gls{COAWST}, we built a domain- and resolution-agnostic \gls{AI} model (\cref{fig:surge_model}) inspired by the U-Net~\cite{Ronneberg2015unet} architecture. The model is an encoder-decoder architecture, with skip connections and fully-connected blocks. It  avoids assumptions about spatial topology or sampling resolution, allowing us to ingest input originating from heterogeneous grids. This approach provides a flexible global mapping between fixed-size input windows and targets. We mitigate overfitting using normalization and dropout.

The encoder-decoder model is trained by minimizing the errors of an $\ell^1$ loss function: $\mathcal{L} = \frac{1}{n} \sum_{i=1}^{n} ||z_i - \hat{z}_i||_1$,
%\begin{equation}\mathcal{L} = \frac{1}{n} %\sum_{i=1}^{n} ||z_i - \hat{z}_i||_1,
%\label{eq:edloss}
%\end{equation}
where $z_i$ is a high resolution ground-truth storm snapshot and $\hat{z}_i$ is the model prediction.  

\subsection{Predicting the environmental effect of interventions with surrogate modeling}\label{sec:img2img}

% \alex{assigned to @alex}

A key challenge in coastal resilience modeling is the lack of data describing how the placement of interventions affect weather impacts in coastal regions. Here we describe how \gls{AI} \glspl{SM} can be used to predict the effect of interventions, given weather data without interventions.

Let $\mathbf{u}(\mathbf{x}, t)$ be a field describing the evolution of one more state variables (e.g., water height, wave direction) on a spatial domain.
%Let $\mathcal{X}$ be a spatial domain containing water and land, and let $\mathbf{u}(\mathbf{x}, t)$ be a field describing the evolution of one more state variables of interest (e.g., water height, wave direction) on $\mathcal{X}$. 
The field starts with an initial condition $\mathbf{u}(\mathbf{x},t=0)$ that is obtained elsewhere (\cref{sec:data}). We can modify the spatial domain by adding in an intervention $I$ (e.g., a layer of oyster reefs near the coast line). Intervention-modified state variables are denoted by $\mathbf{u}_I(\mathbf{x}, t)$, the spatiotemporal field that evolves given the presence of the intervention $I$.

Thus, the effect of an intervention $I$ can be predicted given a method $G_\theta$ for mapping $\mathbf{u}$ to $\mathbf{u}_I$. For a rectangular spatial domain, each $\mathbf{u}(t)$ and $\mathbf{u}_I(t)$ is an image, with each channel giving one of the state variables. Thus, we formulate this task as an image-to-image translation problem, where the \gls{AI} model learns a $G_I$ given a set of input data tuples $(\mathbf{u}(t), \mathbf{u}_I(t), I)$, yielding $\mathbf{u}(t), I \mapsto G_I(\mathbf{u})(t)$.
% the operator:
% \begin{eqnarray}
    % \mathbf{u}(t), \delta &\mapsto& G_\delta(\mathbf{u})(t).
    % \label{eq:neuraloperator}
% \end{eqnarray}
Interventions $I$ are also represented as spatial arrays, and so are added as further channels. We also add in the bathymetry and a binary land-sea mask. Thus, the model input has a number of channels equal to $n_{vars}+n_{interventions}+2$, and its outputs have $n_{vars}$ channels.

The \gls{SM}, given data with $n_t$ time points and $n_\theta$ different interventions, is trained by minimizing
% \begin{equation}
%     \mathcal{L} = \frac{1}{n_t\,n_\theta }\sum_{t,\theta}\mathcal{C}(\mathrm{Crop}(G_\theta(\mathbf{u})(t) - \mathbf{u}_\theta(t))),
%     \label{eq:loss}
% \end{equation}
\begin{equation}
    \mathcal{L} = \frac{1}{n_t\,n_I}\sum_{t,I}||\mathrm{Crop}(G_I(\mathbf{u})(t) - \mathbf{u}_I(t))||^2_F,
    \label{eq:loss}
\end{equation}
where $\mathrm{Crop}$ is a random crop.
%and $\mathcal{C}(Z) = \sqrt{||Z||_F^2 + \epsilon^2}$ is the Charbonnier loss~\cite{lai2018laplacianpyramid}, a differentiable approximation of the $\ell^1$ norm, and $\epsilon = 10^{-3}$. We use Charbonnier loss because it has shown in super-resolution research to be more effective than metrics like squared error at ensuring that models preserve precisely-shaped spatial features. 
Our work uses only a single domain, but the inclusion of bathymetry and the land-sea mask as model inputs enables the model to learn to generalize across geographies.

Because the spatiotemporal fields are driven by multiple physical scales, we use the \gls{SwinIR}~\cite{Liang2021swinir} transformer as the base \gls{AI} model for $G_\theta$. Hyperparameters are given in~\cref{tab:swinir} in~\cref{sec:hyperparameters}. The \gls{SwinIR} model we train from our \gls{COAWST} data has ${\sim}900$k parameters. However, to speed up inference time during the optimization (\cref{sec:optimization}), we use model distillation~\cite{hinton2015distillation} and train a smaller model with ${\sim}100$k parameters (\cref{sec:additionalresults}).

% Because $G_\theta$ maps between function-spaces, we refer to it as a \gls{NO}~\cite{Azizzadenesheli2024neuraloperator}. This formulation is applicable to a wide variety of intervention-prediction problems, depending on the dimensionality and form of the spatial domain $\mathcal{X}$ and the representation of the interventions $\theta$. %~\Cref{fig:surrogate_overview} gives an overview of this approach.

% \begin{figure}[h]
%     \centering
%     \begin{tikzpicture}
%         \randuck
%     \end{tikzpicture}
%     % \includegraphics[width=\linewidth]{figures/surrogate_schematic_v2.png}
%     \label{fig:surrogate_overview}
%     \caption{We summarize our surrogate modeling strategy, showing representative inputs (a sample wave height without an intervention and a sample intervention) and output (the wave height given that intervention).}
% \end{figure}

We randomly sample from the space of oyster interventions (\cref{sec:data}) and use the resulting simulations for model training and validation. Both \gls{SwinIR} models are trained with Adam~\cite{kingma2017adam}.

% \begin{figure}[h]
%     \centering
%     \begin{tikzpicture}
%         \randuck
%     \end{tikzpicture}
%     \caption{schematic of 1d modeling use case (including decomposition of data arrays into diagonals)}
%     \label{fig:1d_usecase}
% \end{figure}

% \begin{figure}[h]
%     \centering
%     \includegraphics{figures/1d_surrogate_data.png}
%     \caption{diagonal decomposition}
%     \label{fig:1d_data_decomposition}
% \end{figure}

\subsection{Intervention optimization}\label{sec:optimization}

Given models of the storm surge and waves impacting a target region, as well as a model of the impact of different intervention configurations, we train an agent to \emph{optimize} the choice of interventions. The goal is to choose a set of interventions that minimize the sum of projected flooding costs, while also considering the costs of implementation and maintenance.  For Tyndall \gls{AFB}, we derive solutions both specifically for Hurricane Michael and across the set of storms described in \cref{sec:generation_results}. 

\subsubsection{Intervention Parameterization}\label{subsec:int_param}
We consider oyster reef (offshore) and sea wall (onshore) interventions. The reefs attenuate waves before they hit the coast, while the sea wall may attenuate waves or redirect water (both storm surge and waves) upon landfall. The agent's action space (\cref{fig:placement_map}) is parameterized as a 63-dimensional vector with continuous and binary values between $0$ and $1$; the first 23 correspond to normalized heights of sea wall segments (up to $5$m) while the last 40 are binary decisions on oyster reef sites.

\subsubsection{Evaluating Flooding Costs}\label{subsec:flooding}

To evaluate flooding cost (Appendix \ref{sec:flood}), we use the combination of a \gls{WOT} model~\cite{eurotop2007, eurotop2018} and gridded cost estimates dependent on the combination of flooding depth and occupancy~\cite{USCorpsEng_2006}. At a high level, we track the flow of water onto land for each hour during a storm, allow its height to equilibrate over the affected gridded area (including bathymetry), and use monetary cost estimates based on insurance claims to compute the overall cost of a storm.  
%A key assumption is that absorption of water will be roughly balanced out by rainfall during a storm. 
The optimization objective / reward $r(I)$ for a set of interventions $I$ over a set of storms $\{j\}$ is the sum of three terms:
\begin{equation}
    r(I) = - CI_I + \sum_j f(j)\left( CS_{0,j} - CS_{I,j}\right),
    \label{reward_eq}
\end{equation}
where $CI_I$ is the cost of installing and maintaining $I$, $CS_0$ refers to the cost of a storm given no interventions, $CS_I$ is the cost of a storm given interventions $I$, and $f(j)$ refers to the expected frequency of storms of the same category (intensity) as $j$ over the lifetime of the interventions. The frequency $f$ was set to $1$ for our analysis of Hurricane Michael, given the rarity of such an event.  Further details of the estimation of rewards are given in Appendices \ref{sec:storm_frequencies}, \ref{sec:damage}, and \ref{sec:cost_int}.

\subsubsection{Continuum-Armed Bandit Problem}\label{subsec:continuous_bandit}
Our intervention selection problem is a one-step decision that is computationally expensive to evaluate and that has a high-dimensional, continuous action space. While our simulation environment is deterministic, it necessarily represents an uncertain future. We therefore frame the problem as a continuum-armed bandit \cite{kleinberg2004continuum} to (1) naturally leverage function approximation to efficiently search our solution space and (2) explicitly incorporate uncertainty, originating both from the environment we model and from the reward estimate in our bandit approach. To solve the problem, we used \gls{STO-BNTS}~\cite{Dai2022sto_bnts}. At each iteration, \gls{STO-BNTS} trains a \gls{NN} so that choosing inputs (actions) that maximize its output is equivalent to sampling from a \gls{GP} posterior for reward, with the \gls{NTK}~\cite{Jacot2018ntk} as the kernel function. To accommodate the high dimensionality of our search space and the high computational expense of obtaining individual samples, we implemented scheduled, prioritized sampling of initial points when optimizing actions~\citep{MartiResendeRibeiro2013} based on the computed reward posterior. Further details are provided in \cref{sec:optimization_alg}.

\section{Results}\label{sec:results}
% In this section we describe a feasibility study that was conducted at APL this summer.  In this feasibility study we explored using a methodology for downscaling low resolution storms from \gls{COAWST}~\cite{Warner2008coawst,warner2010development} to high resolution that would inform the threat modeling method.   We also describe a prototypical surrogate threat model used to model hurricanes and hurricanes with proposed interventions included in the model.  Finally we describe a deep reinforcement learning effort which is used to model, characterize and perform intervention selection.

We evaluated our framework by optimizing the placement and height of sea walls and oyster reefs near Tyndall \gls{AFB} in Florida, which was catastrophically impacted by Hurricane Michael~\cite{beven2019michael}.

\subsection{Storm Surge Results }\label{sec:generation_results}
The encoder-decoder skip connection model (\cref{fig:surge_model}) was trained on low-resolution 5 km \Gls{COAWST} data to learn the relationship between wind and wave fields. It was built to provide a prediction on a single time-step at a time.
%, ensuring full flexibility in terms of both grid size and resolution. 
Our study sought to understand whether the model could be trained to learn this relationship on 5 km data but used on higher resolution data, such as 1 km and 8 m data.

We first collected storm data from \gls{USGS}\footnote{https://www.sciencebase.gov/catalog/item/610acd4fd34ef8d7056893da}, with an hourly temporal resolution and a 5 km horizontal resolution. The data include storms from 2010-2022, over the Gulf, while focusing on dates consistent with periods when storm activity was most significant. Our balanced dataset included an equal number of hurricanes and non-hurricane storms. Data were masked to focus only on points over water and near landfall.  Wind fields, wave fields, and time are the input variables used, and surge (zeta) is used as the target variable.  With a total of 600,000 samples, 400,000 were used for training, of which 80,000 were used for validation, and 200,000 were held out for testing. The model was trained for 60 epochs with early stopping, and used an Adam~\cite{kingma2017adam} optimizer and a ReduceLROnPlateau scheduler.

The model was evaluated on the test set to predict the surge using wind, wave, and time. For normalized predictions, the model's \gls{MSE} was $0.008$, and the \gls{MAE} was $0.073$; unnormalized (meters), the \gls{MSE} was $0.098$, and the \gls{MAE} was $0.234$.
%Given the simplicity of the encoder-decoder skip connection model's architecture, these results are encouraging. 
\cref{fig:predicted_surge_density} (\cref{sec:additionalresults}) shows reasonable distributional agreement between predicted and true surge.

%Using the highly resolved three-way coupled \gls{COAWST} model, Hurricane Michael was simulated, and numerical data was used to evaluate the effectiveness of the surge generator for a storm that has a different grid and resolution.  Michael was specific to the area around Tyndall \gls{AFB}, where Hurricane Michael made landfall, and was generated with an 800 m horizontal resolution.  

\begin{figure}[h]
     \centering
     \includegraphics[width=0.9\linewidth]{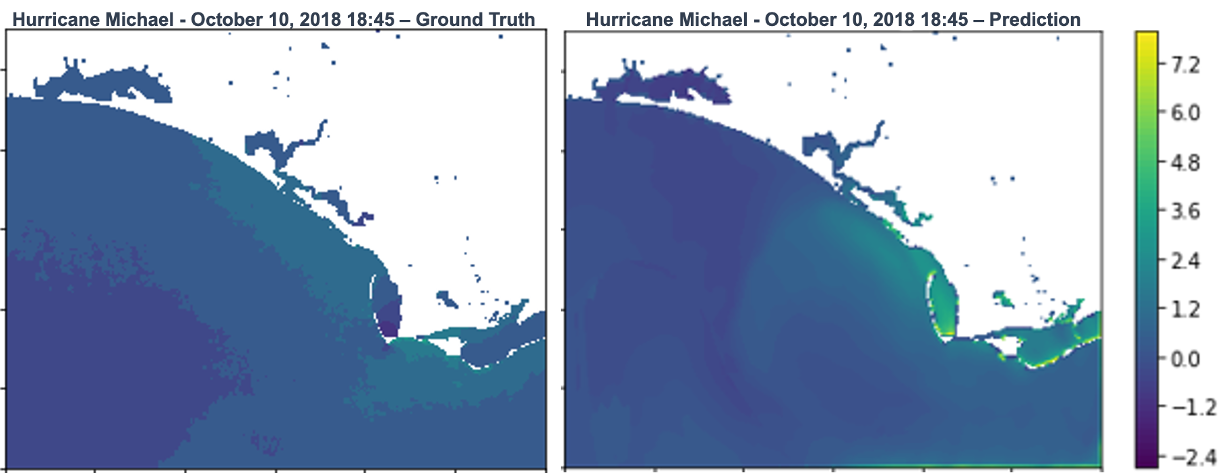}
     \caption{Hurricane Michael Comparing Ground Truth Surge and Predicted Surge.}
     \label{fig:predicted_surge_michael}
 \end{figure}

Next, we evaluated the encoder-decoder skip connection model on Hurricane Michael at an 800 meter horizontal resolution and hourly time resolution.
% When applying the trained encoder-decoder skip connection model to the Michael input wind and wave fields, 
Here, the overall unnormalized \gls{MSE} of the predicted storm surge was $0.03$. The storm surges generated by the numerical model and the predicted storm surge by the encoder-decoder skip connection model are shown in \cref{fig:predicted_surge_michael}.  The magnitude of zeta values is slightly exaggerated, but the values were judged to still be physically consistent.  

We generated twelve storms using \gls{COAWST}: Barry, Delta, Helene, Ian, Ida, Idalia, Irma, Laura, Michael, Nate, Sally, and Zeta. The dates and intensities of the hurricanes are in~\cref{fig:storms} (\cref{sec:hurricanes}). The simulations were run without storm surge to reduce the time complexity of running the three-way coupled \gls{COAWST} model. These storms were created with a horizontal resolution of 1.5 km and an hourly time resolution while covering a broader domain of the entire Gulf. 

% \jared{Can we move Figure 5 to an appendix? we could list the storms generated in the text and save a lot of space}
% \alex{great point!}
% \caption{Gulf of Mexico Storms Generated using COAWST.}

We use a subset of storms in the optimization intervention framework to provide a greater diversity of storms.  By reducing to two-way coupling, the time to generate storms that took on the order of weeks using the three-way coupled model takes now on the order of hours to one day.  This includes using the trained encoder-decoder skip connection model to generate surge for two-way coupled storms.  For a single storm, a 24-hour output from the encoder-decoder skip connection model takes on the order of minutes calling it on a modest CPU and seconds on a A100/H100 GPU.

% Future work will include evaluations for the entire set of storms.  The encoder-decoder skip connection network relies on a significant amount of training data and could benefit from a spatial-temporal convolutional approach.  However, given the goal of this project was to create a model that was specific for the generated COAWST datasets, the current architecture was appealing.

\subsection{Intervention prediction results}\label{sec:surrogate_results}

From the \gls{COAWST} data, we sample ten oyster reefs to use for training and eight for validation. We train \gls{SwinIR} to predict wave direction and wave height and assess with the mean relative error:
\begin{equation}
    \frac{1}{n_t\,n_I}\sum_{t,I}\frac{||G_I(\mathbf{u})(t) - \mathbf{u}_I(t)||_F}{||\mathbf{u}_I(t)||_F}.
    \label{eq:rel_error}
\end{equation}
Interventions primarily change the fields in localized regions, rather than producing significant differences across the entire domain. As such, global metrics like eq.~\ref{eq:rel_error} can fail to fully characterize accuracy. Thus, we compare \gls{SwinIR} to a baseline that assumes that interventions have no effect: 
\begin{equation}
    G_I(\mathbf{u})(t) = \mathbf{u}(t).
    \label{eq:baseline}
\end{equation}

\begin{figure}
    \centering
    \includegraphics[width=0.9\linewidth]{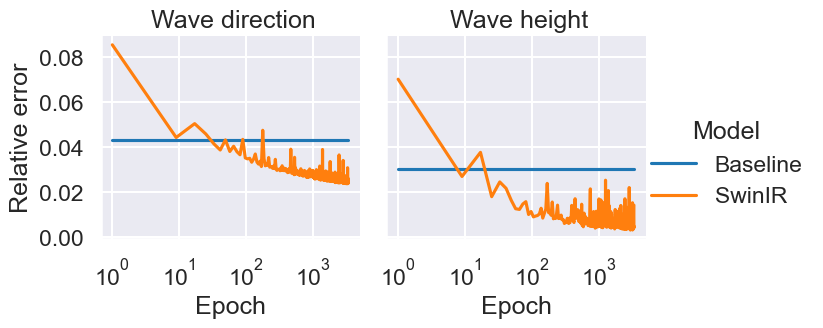}
    \caption{As measured by relative error (eq.~\ref{eq:rel_error}), our \gls{SwinIR} model surpasses the baseline (eq.~\ref{eq:baseline}) in predicting wave fields on our validation cases.}
    \label{fig:surrogate_error}
\end{figure}
\begin{figure}
    \centering
    \includegraphics[width=\linewidth]{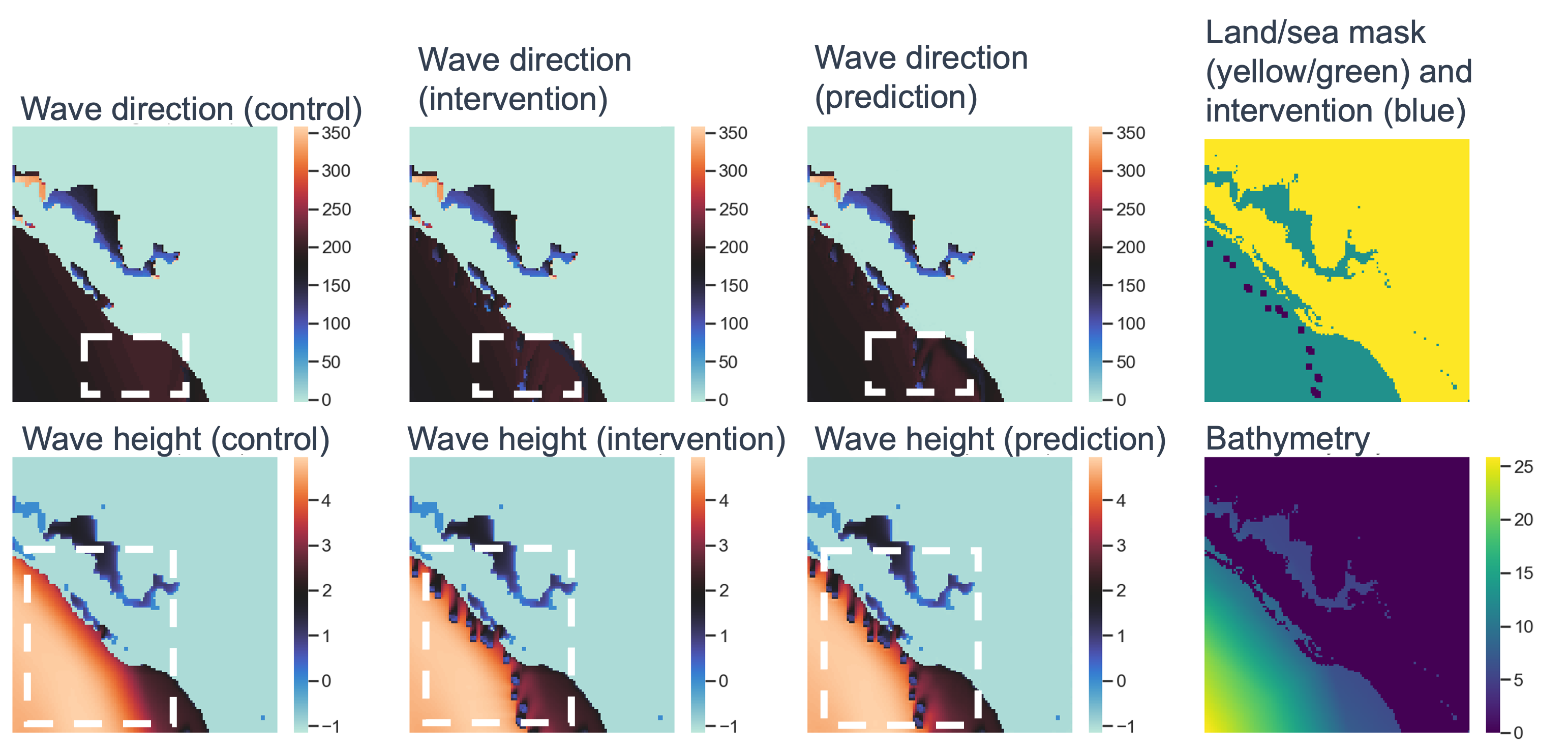}
    \caption{For a given intervention setup and time point from our validation set, we show the intervention-less wave variables (far left); \gls{COAWST}-predicted variables (middle left); \gls{SwinIR}-predicted variables (middle right); and land-sea mask, intervention locations, and bathymetry (far right). For both wave direction and height, \gls{SwinIR} captures how the oyster reefs change the waves.}
    \label{fig:surrogate_results}
\end{figure}

In~\cref{fig:surrogate_error}, we show that the \gls{SwinIR} model can attain a relative error of ${\sim}2\%$ for predicting wave direction and ${\sim}1\%$ for predicting wave height. For both state variables, this is approximately half the error as predicted by our baseline. In~\cref{fig:surrogate_results}, we further visualize \gls{SwinIR} performance and show how the model accurately predicts how a given oyster reef reduces wave height. 

\begin{figure}
\includegraphics[width=0.49\linewidth]{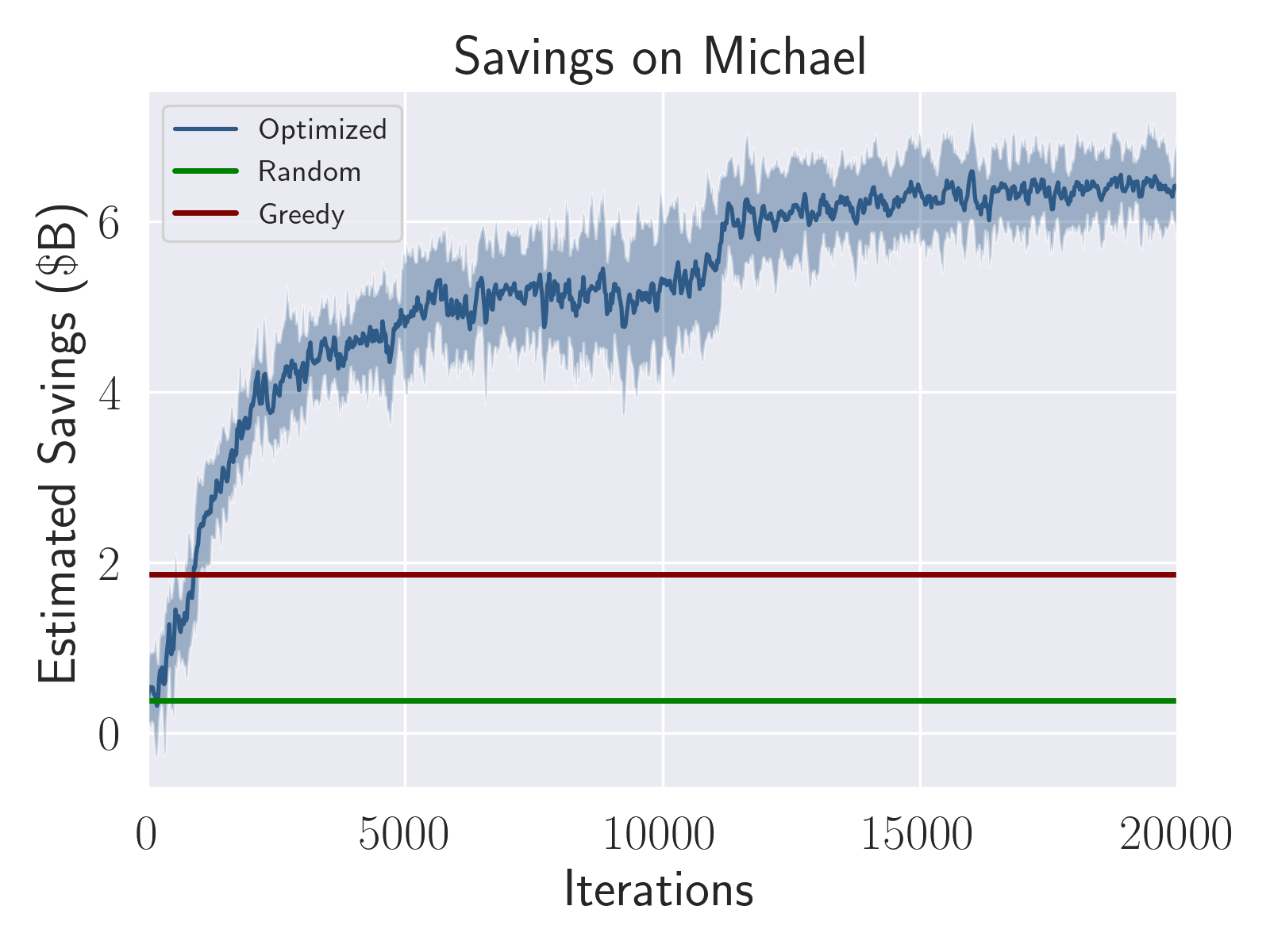}
\includegraphics[width=0.49\linewidth]{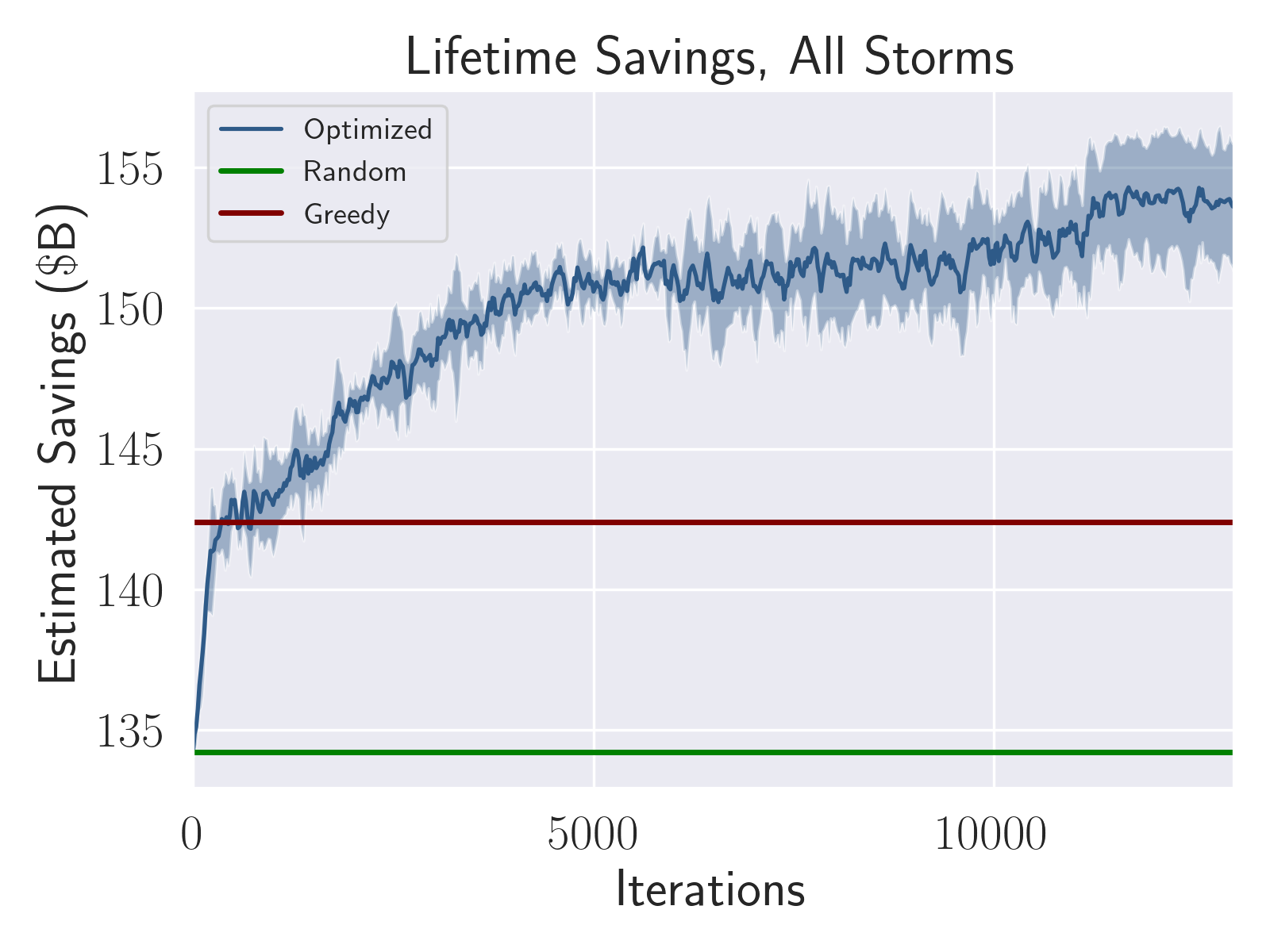}
\caption{Progression of rewards (Eq.~\ref{reward_eq}) in intervention optimization with \gls{STO-BNTS} compared with randomly chosen and greedy intervention choices. Greedy lines deploy all possible interventions up to their maximum height, while random lines use randomized actions with an untrained network. Each ``optimized'' iteration refers to the training of an \gls{NN} reward posterior and action selection based on its maximization.  Left: Training for Hurricane Michael only. Shading for the optimized curve reflects 5 random seeds.  Right: Training for our distribution of storms and integrating over the planned 50-year lifespan of the interventions. Shading reflects 4 random seeds.
}
\label{fig:learning_curves}
\end{figure}

\subsection{Intervention Optimization results}\label{sec:optimization_results}
Optimized interventions were derived for two cases: Hurricane Michael only and for the combination of Hurricanes Barry, Ian, Idalia, Laura, Michael, Nate, and Sally. Estimated savings of selected interventions are displayed in~\cref{fig:learning_curves}. 
%The flooding reduction is clipped for visualization purposes.

For Hurricane Michael, we found that randomized interventions would have amounted to little savings- on average, the savings provided by such choices are roughly balanced by the costs of the interventions.  The greedy solution is an improvement, but does not nearly match the roughly \$6B in net savings that interventions optimized for that particular storm could have allowed. 

When optimizing against the distribution of storms, we consider the problem more practically. We seek interventions that protect high-value areas from flooding as effectively as possible, given a set of storms designed to be representative of possible future occurrences. To properly weigh costs, we multiply each storm by its relative frequency (eq.~\ref{reward_eq}) and (roughly) cover the different relevant categories of storms- from consequential but small storms (less than Category 1) to a direct hit from a Category 5 hurricane.  We sum over the occurrences of all types over the nominal 50-year \footnote{This is conservative: properly maintained sea walls and oyster reefs may last significantly longer.} lifespan of interventions to arrive at the large numbers in the right panel of \cref{fig:learning_curves}.  Here, even randomized interventions from our set of options lead to big savings (note that the cost of interventions is counted only once). However, we find that large gains are still available through intervention optimization.

\begin{figure}
\includegraphics[width=0.329\linewidth]{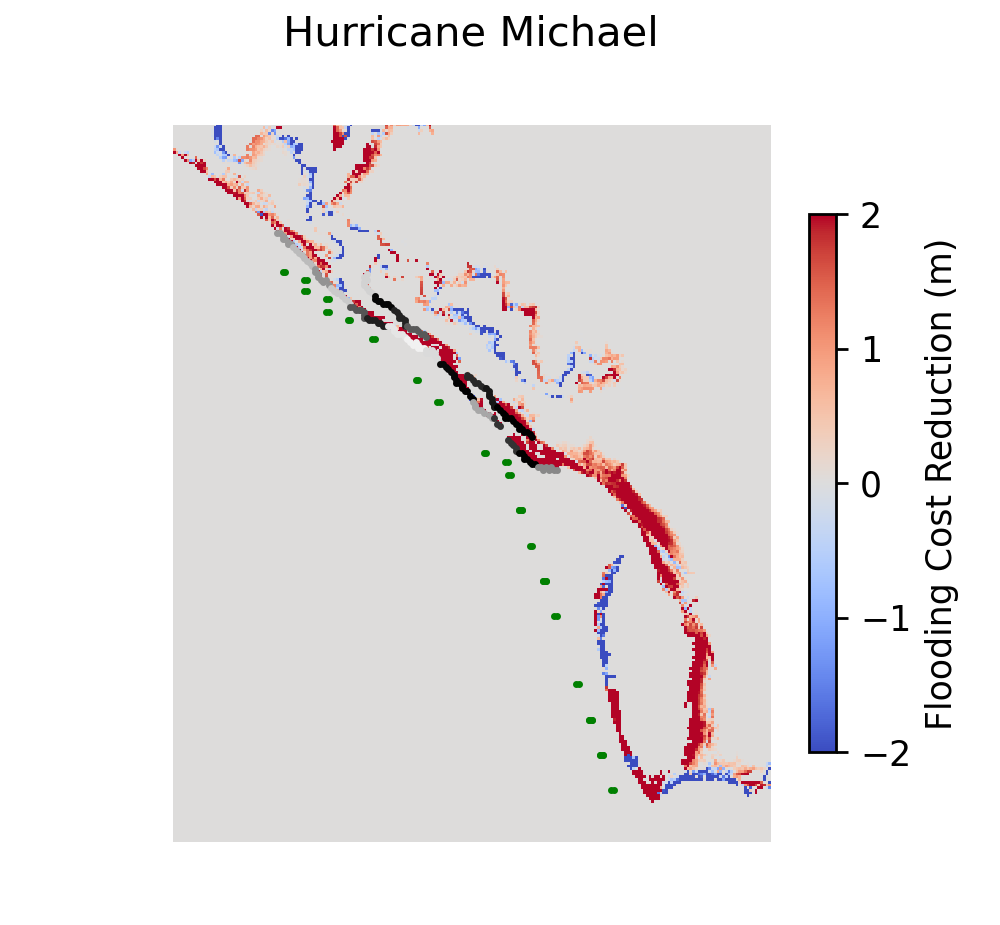}
\hfill
\includegraphics[width=0.329\linewidth]{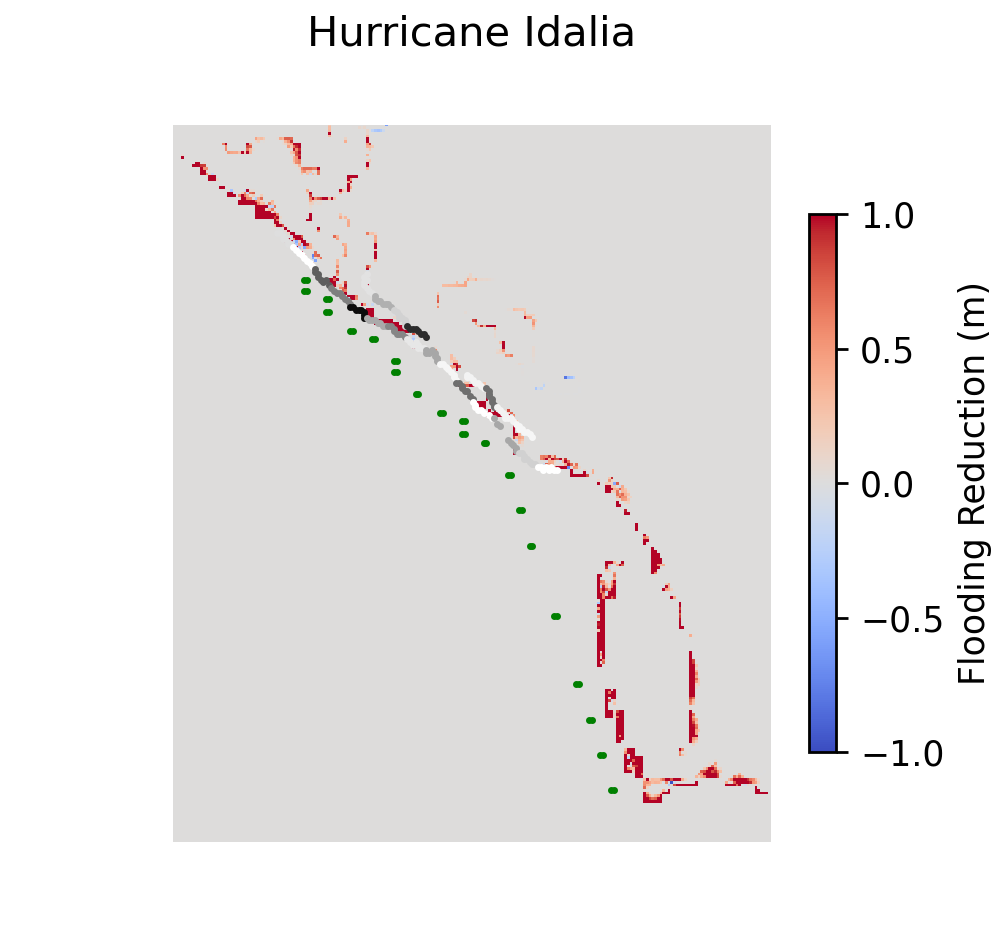}
\hfill
\includegraphics[width=0.329\linewidth]{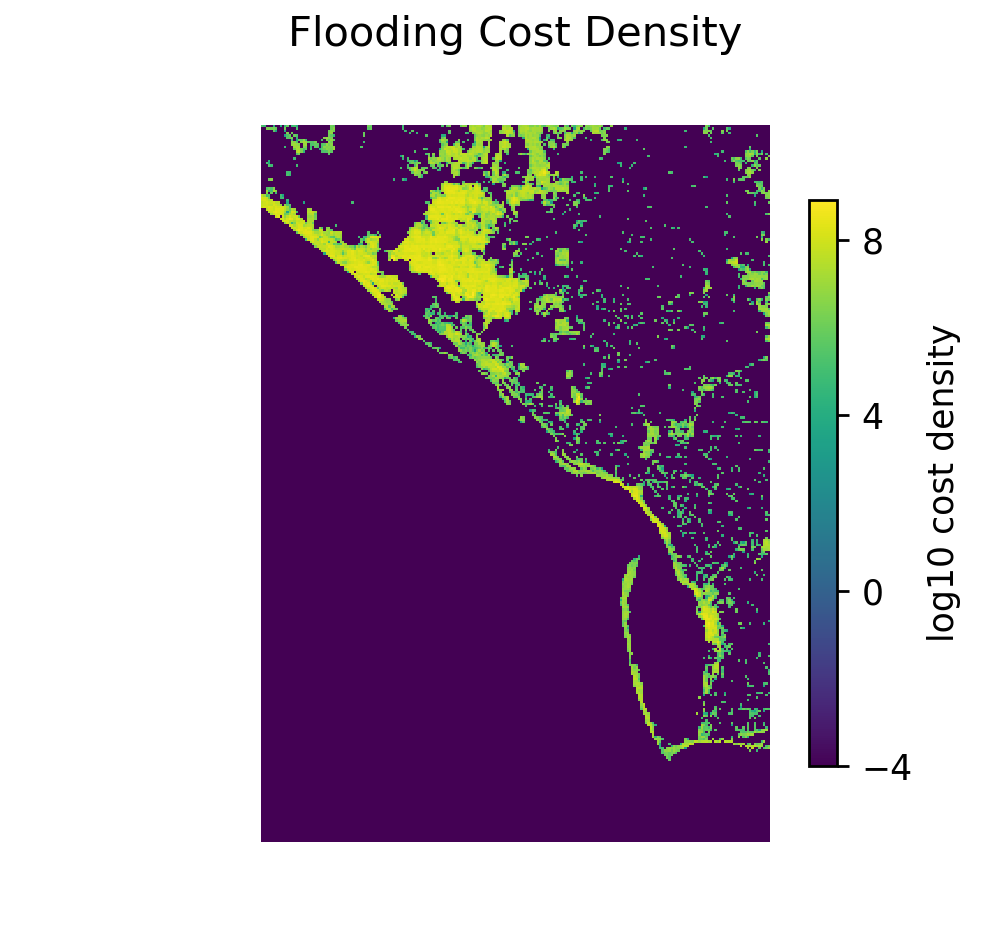}
\caption{Optimized interventions. Green dots represent prescribed oyster reef sites; white, grey, and black dots refer to the height of sea wall deployed at a potential site (dark is higher, up to 5m).  Red/blue shows where interventions reduce/increase flooding relative to no interventions. The flooding reduction is clipped for visualization purposes. Left: Training only on Hurricane Michael.  Center: Interventions on Hurricane Idalia, based on training over all storms.  Right: Cost mask.}
\label{fig:placement_map}
\end{figure}

\Cref{fig:placement_map} provides a visualization of flooding reduction based on optimized interventions, both for the Michael-only optimization (left) and for Hurricane Idalia (center) in the full optimization.  The cost grid that guides the optimization is shown on the right.  We may observe that the Michael-only optimization concentrates resources near the center of the image, consistent with the storm track (\Cref{fig:predicted_surge_michael}). The full optimization concentrates resources more toward the area with high cost density.  Both optimizations deploy oyster reefs to protect the peninsula in the bottom right of the images, where a sea wall would be impractical. Additional flooding reduction maps are given in Appendix \ref{sec:more_interventions}.

\section{Discussion}\label{sec:discussion}

We have developed a framework (\cref{fig:schematic}) that characterizes and optimizes current coastal resilience capabilities. % while guiding development of new intervention types based on identified gaps. 
Our approach combines fast storm data generation, effect prediction, and optimization over potential interventions. When used to choose oyster reefs and seawalls near Tyndall \gls{AFB}, our models suggest that optimization could potentially save tens of billions of dollars in flooding damage. 
%Our results are obtained using \gls{AI} models that are significantly faster than traditional numerical computations.

Our results consider the area surrounding Tyndall \gls{AFB}; however, we see our framework as being extensible to a wide range of geographies, storms, and intervention types. The accuracy and data efficiency of our models could be enhanced by the inclusion of other architecture components, such as spatiotemporal convolutions (e.g.,~\cite{li2021fourier}), diffusion-based forecast refinement~\cite{Lippe2023refiner}, or physics-based regularization~\cite{collins2023rapid}. More generally, future work could explore the integration of other data sources, including remote sensing, into our %prediction and optimization 
pipeline. 
%This could allow for more reliable damage estimates and provide additional data for fine-tuning models. 
The impact of different assumptions on future storms -- including their frequency, intensity, and trajectory-- on intervention optimization could additionally be explored, perhaps through consideration of a sequential decision-making process, as in~\cite{Bhattacharya2025}.

%\alex{@jared can you make some comment about distributional assumptions of storms and how that affects optimal interventions? we can point to~\cite{Bhattacharya2025} as an example approach. this also ties into maintenance costs for interventions}

\clearpage

\section*{Acknowledgments}

This work was supported by internal research and development funding from the Research and Exploratory Development Mission Area of the Johns Hopkins Applied Physics Laboratory. Thanks to Jennifer Boothby, Sarah Herman, Marisa Hughes, Heather Hunter, Christine Piatko, Elizabeth Reilly, and Kristen Ryan for help refining the concept of our framework.

\bibliographystyle{unsrt}
\bibliography{references}

%%%%%%%%%%%%%%%%%%%%%%%%%%%%%%%%%%%%%%%%%%%%%%%%%%%%%%%%%%%%

\appendix

\glsresetall

\section{Additional technical details}

\subsection{Generating data with COAWST}\label{sec:coawst_extras}

\Gls{COAWST}~\cite{Warner2008coawst,warner2010development} couples a set of models in order to capture the complex dynamics of coastal and marine environments. It includes 1.) an ocean model (\gls{ROMS}~\cite{Shchepetkin2005roms}) that simulates the physical state of the ocean, including temperature, salinity, currents, and sea level, 2.) an atmospheric model (\gls{WRF}~\cite{Skamarock2019wrf}) that represents atmospheric processes, including wind, temperature, pressure and humidity, 3.) a wave model (\gls{SWAN}~\cite{booij1999swan,swan2020manual}) that simulates the propagation and transformation of ocean waves, and 4.) a sediment model (\gls{USGS} Community Sediment
Modeling System) that simulates the movement of sediment within the water column and across the seabed, influenced by currents and waves.

When we recreated Hurricane Michael in \gls{COAWST}, we did so at an 800 m resolution in the \gls{WRF} domain and used a horizontal grid resolution of approximately 300 m for the \gls{SWAN} and \gls{ROMS} domain. 

\subsection{Hurricanes of interest}\label{sec:hurricanes}

 \begin{figure}[h]
     \centering
     \includegraphics[width=0.6\linewidth]{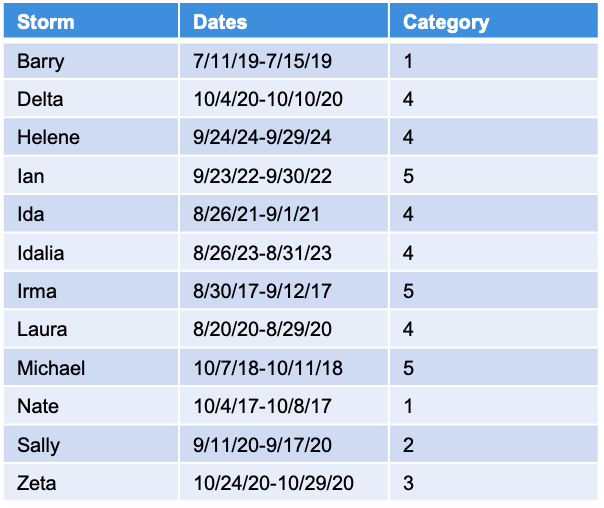}
     \caption{Gulf Storms Listed By Name, Time of Landfall, and Category. These storms were generated using a two-way coupled model and surge (zeta) was generated for a subset of storms using the Encoder Decoder Skip Connection model.}
     \label{fig:storms}
 \end{figure}

\Cref{fig:storms} lists the storms used in our study by name, time of landfall, and category.

\subsection{Hyperparameters}\label{sec:hyperparameters}

\Cref{tab:swinir} gives hyperparameters used to train the \gls{SwinIR} models (\cref{sec:img2img}, \cref{sec:surrogate_results}). 
%\Cref{tab:nts} gives hyperparameters used to deploy the \gls{NTS} results (\cref{sec:optimization}, \cref{sec:optimization_results}).

\begin{table}[h]
    \centering
    \begin{tabular}{c|c}
    Hyperparameter                          &   Value\\\hline
    Learning rate                           &   $10^{-3}$ \\
    Number of training epochs               &   $2000$\\
    Batch size                              &   $16$ \\
    Crop size                               &   $128 \times 128$\\
    Window size                             &   8\\
    Depths (Full model)                     &   6, 6, 6, 6\\
    Depths (Distilled model)                &   6\\
    Embedding dimension (Full model)        &   60\\
    Embedding dimension (Distilled model)   &   36\\
    Number of heads (Full model)            &   6, 6, 6, 6\\
    Number of heads (Distilled model)       &   6\\
    MLP ratio                               &   2
    \end{tabular}
    \caption{Hyperparameters used to set up and train the \gls{SwinIR} model. We adapted the official implementation~\cite{swinir2021github}; the documentation explains each variable's meaning. See~\cref{sec:img2img} for explanations of the full vs. distilled model.}
    \label{tab:swinir}
\end{table}

\section{Additional technical results}\label{sec:additionalresults}

The \gls{SwinIR} model we train from our \gls{COAWST} data has ${\sim}900$k parameters. However, to speed up inference time during the optimization (\cref{sec:optimization}), we use model distillation~\cite{hinton2015distillation} and train a smaller model with ${\sim}100$k parameters (\cref{sec:additionalresults}), using the predictions of the larger model as targets:
\begin{equation}
    \mathcal{L}_{distillation} = \frac{1}{n_t\,n_I}\sum_{t,I}||(\mathrm{Crop}(G_{I,large}(\mathbf{u})(t) - G_{I,small}(\mathbf{u})(t))||_F^2.
    \label{eq:distillation}
\end{equation}

\begin{figure}
    \centering
    \includegraphics[width=0.9\linewidth]{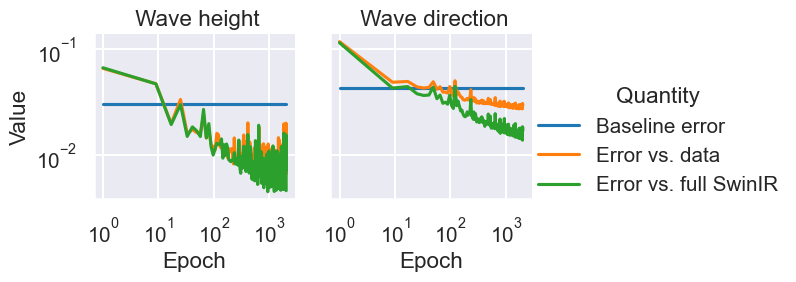}
    \caption{As measured by relative error (eq.~\ref{eq:rel_error}), model distillation lets a more lightweight \gls{SwinIR} model retain accuracy in predicting wave height while degrading only somewhat in predicting wave direction. ``Baseline error'' is the error of the prediction baseline (eq.~\ref{eq:baseline}), ``Error vs. data'' is the error of the smaller \gls{SwinIR} model vs. the \gls{COAWST} data, and ``Error vs. full SwinIR'' is the error of the smaller \gls{SwinIR} model vs. the full \gls{SwinIR} model.
    }
    \label{fig:distillation_error}
\end{figure}

In~\cref{fig:distillation_error}, we additionally show that the model distillation procedure~\cite{hinton2015distillation} we use to obtain a faster-to-evaluate \gls{SwinIR} model retains accuracy and still surpasses the baseline.

\begin{figure}
     \centering
     \includegraphics[width=0.7\linewidth]
     {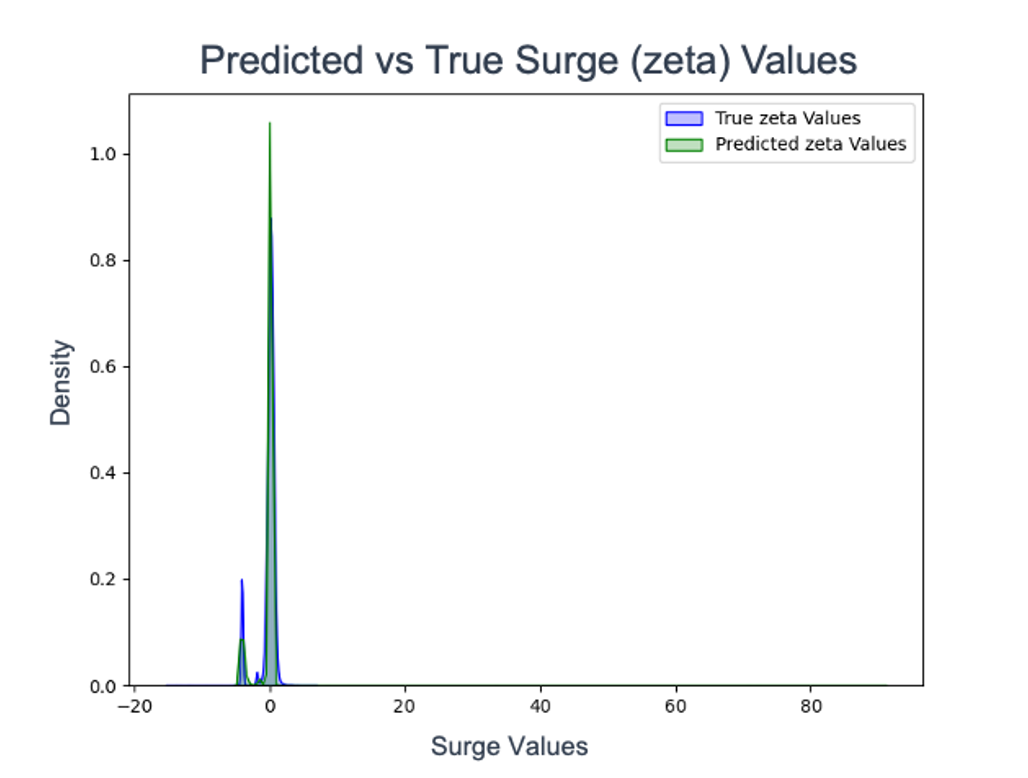}
     \caption{Density Plot Showing Distributional Agreement between AI-Generated Surge (zeta) and Ground Truth Surge (zeta) for the 5 KM \gls{COAWST} test data set.}
     \label{fig:predicted_surge_density}
 \end{figure}

In~\cref{fig:predicted_surge_density}, we visualize the performance of the surge prediction model (\cref{fig:surge_model}), showing that the distribution of true and predicted surge values are comparable.

\section{Further details on interventions, flooding, and damage cost assessments}\label{sec:general_interventions}

\subsection{Storms and Frequencies}\label{sec:storm_frequencies}
For intervention optimization, we considered two scenarios: optimization for Hurricane Michael specifically and optimization against a set of storms.  This included AI-generated surge estimates for Hurricanes Barry, Ian, Idalia, Laura, Michael, Nate, and Sally.  Note that, while Ian was a Category 5 Hurricane, it did not impact Tyndall AFB as significantly as the other storms considered and therefore is taken as a $<1$ Category storm here. The relative frequencies $f$ of these storms in Equation \ref{reward_eq} were $[3, 5, 2, 2, 1, 3, 3]$, respectively.  These numbers represent approximate recurrences over a projected 50-year lifetime of the interventions.  Given the spread in the categories of the storms, we project their summation to approximately reflect the total expected storm occurrences in the region while the interventions in question would be deployed.  That is, we estimate roughly one Category 5 hurricane (represented by Michael), four Category 3-4 hurricanes (represented by Idalia and Laura), six Category 2 hurricanes, and eight significant Category 1 and under storms. These numbers are necessarily imprecise, but reflect historical trends \citep{NOAAHRD_USHurricanes_2024}.  It should also be noted that the impact of climate change and the unknown lifetime of interventions (which may well exceed 50 years) will introduce uncertainty into these estimates.

\subsection{Computing flooding and flooding damage}\label{sec:flood}
%\alex{@jared can you go through this section and make sure it's representative of the final flooding model?}
%\jared{checked the equations- this is what we're running}
A (pre-existing) \gls{WOT} model~\cite{eurotop2007,eurotop2018} was used to quantify flooding over the affected region. This set of equations computes the volume of water flowing onto land based on the land bathmetry (height) plus any seawall barrier, the \gls{SWL}, and the significant wave height ($H_{sig}$). The \gls{WOT} model is documented in the EurOtop Wave Overtopping Manual~\cite{eurotop2007}; an example of its use is given by Suh \& Lee~\cite{suhlee2023typhoon}. A key assumption of this analysis is that any absorption of water during the storm is balanced by additional rainfall.

%In that paper, the WOT is used as a boundary condition on the terrestrial side of the system.
The principal formula used for \gls{WOT} is
    \begin{equation}
         \frac{q}{\sqrt{gH^3_{m0}}}= a \exp(-(bR_c/H_{m0})^c),
    \end{equation}
where $q$ is volume of water in $m^2/s$, $g$ is acceleration due to gravity, $H_{m0}$ is the significant wave height (the spectral moment of wave height), $a$ is the scale parameter, $b$ is the shape parameter, and $R_c$ is the freeboard (the difference between the land and wall height and the still water level). The breaker parameter $\xi$ is introduced for more specific use cases of this equation \cite{eurotop2018}. For this demonstration, we chose $a = 1, b = 0.75, c = 1$, and $\xi = 1$, which fall within the ranges described in \cite{eurotop2007} and \cite{eurotop2018}.
%\alex{are $a,b,c,\xi$ all dimensionless?}
%\jared{yes}
%\alex{can we say a bit more about what's meant by ``freeboard''?}
%\jared{it's mentioned above now (someone else put it in, think it is clear)}
Rearranging the above equation to solve for water volume $q$ gives
\begin{equation}
q = \sqrt{gH^3_{m0}} \cdot a \exp\left(-(bR_c/H_{m0})^c\right)
\end{equation}
This formula works for positive and zero freeboard. For negative freeboard,
\begin{equation}
q = 0.6 \cdot \sqrt{g\cdot|R_c^3|} +0.0537\cdot\xi\cdot\sqrt{gH^3_{m0}}
\end{equation}
% \begin{figure}[htb]
% \centering
% \begin{tikzpicture}
%         \randuck
%     \end{tikzpicture}
% % \includegraphics[width=.7\linewidth]{figures/overtopping_diagram.png}
% \caption{\label{fig:overtopping_diagram} \Gls{WOT} and overflow for positive, zero and negative freeboard (reproduced from \cite{eurotop2007}).}
% \end{figure}

In addition to water flowing from ocean to land, we account for the possibility that water flows back to the ocean from saturated grid cells along the coast. For a given cell on the land-sea boundary, we define $R_{c, out}$ as the difference between the maximum of its bathymetry plus the seawall height (if present) and the local water level and the sum of its bathymetry and the local flooding.  If $R_{c, out} <0$, water flows back into the ocean according to
\begin{equation}
    q = 0.6 \cdot \sqrt{g\cdot|R_{c, out}^3|}.
\end{equation}
In this \gls{WOT} method, as water comes onto land during a storm, that volume of water is spread out to low-lying adjacent areas. Our solver ensures that water flows `downhill', such that the height in one gridcell is reduced by a flux into the adjacent cell until the water heights equalize when accounting for the topographic heights. The water volume flowing onto land and spread of all water is updated every hour for the duration of the storm.

%\alex{we probably need to say a bit more here? e.g., timesteps for evolving the flooding?}
%\jared{i added the time steps, not really sure more detail is needed (plus i still don't understand why we're re-computing things with smaller tolerances}

% In past coastal resilience work at APL, flooding was estimated by comparing \gls{SWL} and topographic (land) height, such that all land below \gls{SWL} was `flooded'. This is an over-estimate of the flooding attributable to high SWL, or storm surge, as water cannot flow infinitely onward due to limitations in volume and energy and it also cannot flow over intervening high areas to reach inland low areas. This prior method also neglects the effects of waves on flooding completely. Now, as water comes onto land during a storm, that volume of water is spread out to low-lying adjacent areas. Our solver ensures that water flows `downhill', such that the height in one gridcell is reduced by a flux into the adjacent cell until the water heights equalize when accounting for the topographic heights.

% \jared{I think the following section should be removed, once we have final results.  These solutions were not in the set we considered and the flooding code changed after these plots were made.}
% \alex{donezo: moved to `baseline.tex`}

\subsection{Estimating Flooding Damage}\label{sec:damage}
\begin{figure}
    \centering
    \includegraphics[width=0.49\linewidth]{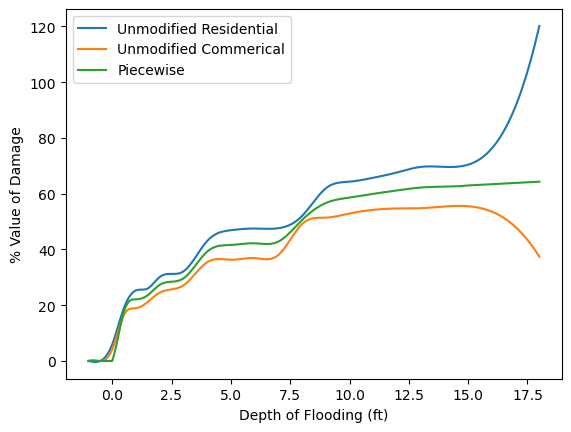}
    \includegraphics[width=0.49\linewidth]{figures/bandit_results/cost_mask.png}
    \caption{Left: splines of the depth to damage estimates in \cite{USCorpsEng_2006}. Right: cost map used to compute the total cost of flood damage; colors reflect $\log10$ scale of cost density.}
    \label{fig:costs_grid}
\end{figure}

%\alex{@jared can you check and make sure this is accurate}
%\jared{i updated}
Flooding estimates are translated to monetary costs using a grid-based strategy. At each timestep, the estimated depth of flooding $d$ for a given land cell is used to estimate a percentage of damage inflicted on all structures in that cell. The product of this damage percentage $D$ and the total value of structures in a cell $v$ is taken as the estimated cost of flooding in that cell. The storm cost $CS$ that contributes to the optimization objective \ref{reward_eq} is the negative summation of these estimates across all land cells $i$, defined by
\begin{align}
    CS(t) = -\sum_{i} d_{i}(D_{i}(t)) \cdot v_{i}.\label{eqn:reward}
\end{align}
Long-term flooding depth to percent damage tables from the US Army Corps of Engineers for residential and commercial structures are averaged and splined to provide a continuous function mapping the amount of flooding in a given cell to the percentage of damage caused to all structures in that cell \cite{USCorpsEng_2006}. The splined values are shown in Table \ref{tab:depth_percent_damage}. In these estimates, some damage to residential structures is assumed even with $0$ feet of flooding, due to their basements. Although accurate, this implies that even with perfect intervention, the agent will receive some negative reward. In addition, the cubic growth of the splines can lead to extreme swings near the boundaries. To mitigate these issues, the percent damaged by $0$ ft of flooding is set to zero and the following C0-continuous, piecewise depth to damage function is used:
\begin{align}
    d(D) = \begin{cases}
    0 & D < 0\\
    S(D) & 0\le D < 15\\
    100\cdot \left[S(15) - S(14.99)\right] \cdot (D-15) + S(15) & D\ge15.
    \end{cases}\label{eqn:depth2damge}
\end{align}
Here $S$ is the spline of Table \ref{tab:depth_percent_damage} and the third portion of the piecewise function is a finite difference approximation of a linear mapping of $d$ to $D$ at the end of the spline. \Cref{fig:costs_grid} (left) displays eq.~\ref{eqn:depth2damge}, as well as the splines for the unmodified residential and commercial depth-to-damage estimates.

Per-cell monetary value estimates $v$ are obtained by using built-up surface data from the \gls{GHSL} project \cite{GHS_V, GHS_S} as well as cost of construction estimates from the Coldwell Banker Richard Ellis Group~\cite{CBRE_US_2024}, a real estate investment firm that conducts research into real estate trends. GHSL data is composed of 100 m grid cells specifying the gross building height $GBV$ and gross building surface area $GBA$. The average building height $ABH$ in a given cell $j$ is defined by $ABH_j = \frac{GBV_j}{GBA_j}$. An average floor height of $3.5$ meters is used to translate these heights to floors per the total area of buildings in that cell $ANF_j$. Finally, the total value of buildings including multiple floors $TV_j$ is computed as
\begin{align}
    TV_j = ANF_j \cdot GBA_j \cdot \$ 3552.09,
\end{align}
where $\$3552.09$ is the cost of construction per square meter.
This cost grid is resampled to the larger land grid by using accumulating costs at nearest land cell centroids which provides $v_i$ in Equation \eqref{eqn:reward}, which is shown in \cref{fig:costs_grid}.
% \begin{figure}
%     \centering
    
%     \caption{}
%     \label{fig:depth2damage}
% \end{figure}

%\jared{I don't know why the NEEDS TO BE REPLACED comment is there.}
%\alex{I don't either, it was in the original DID report}
%\jared{I just removed it.}

\begin{table}[h!]

    \centering
    \begin{tabular}{|c|c|}
        \hline
        \textbf{Depth (ft)} & \textbf{Damage (\%)} \\
        \hline
        \hline    
        -1.0  & 0.00  \\
        -0.5  & 0.00  \\
        0.0   & 0.00*  \\
        0.5   & 17.85 \\
        1.0   & 22.10 \\
        1.5   & 23.60 \\
        2.0   & 27.30 \\
        3.0   & 29.60 \\
        4.0   & 39.20 \\
        5.0   & 41.60 \\
        6.0   & 42.20 \\
        7.0   & 42.80 \\
        8.0   & 50.55 \\
        9.0   & 56.65 \\
        10.0  & 58.60 \\
        11.0  & 59.95 \\
        12.0  & 61.15 \\
        13.0  & 62.20 \\
        14.0  & 62.50 \\
        15.0  & 62.95 \\
        \hline
    \end{tabular}

    \caption{Depth-to-damage values used for splining. These values were obtained from \cite{USCorpsEng_2006} by averaging the commerical and residential damage estimates. In addition, the damage at $0$ft of flooding is enforced to be zero.}
    \label{tab:depth_percent_damage}
\end{table}

\subsection{Estimating the Cost of Interventions}\label{sec:cost_int}

While additional interventions are possible, this study considers concrete seawalls and oyster reefs. Seawall installation was priced as a function of length and height, with a maximum height of 5m being considered.  The particular function used was

\begin{equation}
C_\text{seawall}(l) = \$2050l H^{1.3},
\end{equation}

where $l$ is seawall length in meters and $H$ is the height of the segment in question.  This expression was derived from a log-log fit of recent sea wall installation data in Louisiana \cite{usaceChalmette2009,cpraLafitte2024,cpraRosethorne2022,cpraLafittePress2020}. Maintenance was priced at $1\%$ of the upfront cost annually for the 50-year expected lifetime of the wall (that is, it multiplied the above estimate by $1.5$.  The maximum total cost of all sea wall options was $\$1.75$B.

Oyster reef installations on the scale of that proposed here are thus far rare.  Our cost estimates are based on a recent restoration project in the Chesapeake Bay \citep{oysterUpdate2024}, which cost roughly $\$65k$ per acre.  Engineering, permitting, and maintenance were lumped into an extra $20\%$ fee. The total cost of all oyster reef installations considered was $\$494$M.

\section{Intervention Optimization Strategy}\label{sec:optimization_alg}

As mentioned in Section \ref{subsec:continuous_bandit}, we chose to frame our optimization problem as a continuum-armed bandit~\cite{kleinberg2004continuum}. We applied \gls{STO-BNTS} to solve the problem, modifying the neural-network-based approach in~\cite{Dai2022sto_bnts} slightly. In particular, we implemented scheduled, prioritized sampling of initial points in action selection, allowing the optimizer to then optimize these initial choices using the Adam optimizer \cite{kingma2017adam}.  Each action taken by the agent (after initial data collection) was the best result of optimization of $100$ initial actions, with some of those starting points being taken from an ``elite'' population stored in the replay buffer. These elites represent the top $5-10\%$ of evaluated actions up to the time in question. The frequency of elite starting point selection ranged from $10\%$ early in training to $90\%$ late in training, allowing the agent to explore more early and then increasingly focus on advantageous actions.  Note that this prioritization did not apply to fitting the neural network, nor did it force the agent to repeat elite actions from the replay buffer.

The full algorithm employed is given in Algorithm \ref{alg:stobnts} and the hyperparameters used are given in Table \ref{table:stobnts_hyper}.

\begin{algorithm}[t]
\caption{Sample-Then-Optimize Batch Neural Thompson Sampling~\citep{Dai2022sto_bnts} with Prioritization}\label{alg:stobnts}
\begin{algorithmic}[1]
\For{$t = 1,2,\dots,T$}
    \State Construct NN $f(x;\theta)$ and multiply its output by $\beta_t$
    \For{$i = 1,2,\dots,B$}
        \State Sample $\theta_{0} \sim \mathrm{init}(\cdot)$
        \State Sample $\theta'_{0} \sim \mathrm{init}(\cdot)$ and set the parameters of $\theta'_{0}$ in the last layer to $0$
        \State Set $f_t^{i}(x;\theta)=f(x;\theta_{0})+\langle\nabla_{\!\theta}f(x;\theta_{0}),\theta'_{0}\rangle$
        \State Use observation history $\mathcal{D}_{t-1}$ to train $f_t^{i}(x;\theta)$ with the loss
        \[
\mathcal{L}_{t}\bigl(\theta,\mathcal{D}_{t-1}\bigr)
  = \sum_{\tau=1}^{t-1}\sum_{j=1}^{B}
    \bigl( y_{\tau}^{\,j} - f_{t}^{\,i}(\mathbf{x}_{\tau}^{\,j};\theta) \bigr)^{2}
    + \beta_{t}^{2}\sigma^{2}\,\lVert \theta - \theta_{0} \rVert_{2}^{2}
\]
        \Statex\hspace{\algorithmicindent}\hspace{\algorithmicindent}(initialize with $\theta_{0}$ and run gradient descent with minibatch size $M$)
        \State $\theta_{t}^{i} \gets \arg\min_{\theta}\, \mathcal{L}_t(\theta,\mathcal{D}_{t-1})$
        \State Choose $\mathbf{x}_{t}^{i} \gets \arg\max_{x\in\mathcal{X}} f_t^{i}(x;\theta_{t}^{i})$, where arg max is found via Adam with restarts, 
        \Statex\hspace{\algorithmicindent}\hspace{\algorithmicindent}initial points are sampled according to prioritization schedule 
    \EndFor
    \State Query the batch $\{\mathbf{x}_{t}^{i}\}_{i=1,\dots,B}$ to obtain
           $\{y_{t}^{i}\}_{i=1,\dots,B}$ and add them to $\mathcal{D}_{t-1}$
\EndFor
\end{algorithmic}
\end{algorithm}

\begin{table}[h]
    \centering
    \begin{tabular}{c|c}
    Hyperparameter                          &   Value\\\hline
    Initial points sampled                  &   1,000\\
    Number of iterations, $T$                 &   20,000\\
    Batch size, $B$                           &    1\\
    Prediction weight $\beta_t$                                   & 1 \\
    Observation noise variance $\sigma^2$ & $10^{-2}$\\ \hline
    Reward network hidden layer widths      & [128, 128] \\   
    Reward network activation               & $\mathrm{ReLU}$ \\
    Reward network weight std & $1.5/\sqrt{128}$  \\
    Reward network bias std & 0.05 \\
    Reward Learning rate                           &   $10^{-3}$ \\
    Minibatch size, $M$ & 128 \\
    Epochs per reward training & 20\\
    Max reward training steps & 1000 \\ \hline
    Elite fraction                          & 0.05 (Michael), 0.1, (All storms)\\
    Prioritization Schedule                 & $\min(0.00008 * t, 0.9)$\\
    Action learning rate & $10^{-3}$\\
    Action learning steps & 50\\
    \end{tabular}
    \caption{Hyperparameters used for Sample-Then-Optimize Batch Neural Thompson Sampling.}
    \label{table:stobnts_hyper}
\end{table}

\newpage

\section{Additional Optimized Interventions}\label{sec:more_interventions}
\begin{figure}
\includegraphics[width=0.325\linewidth]{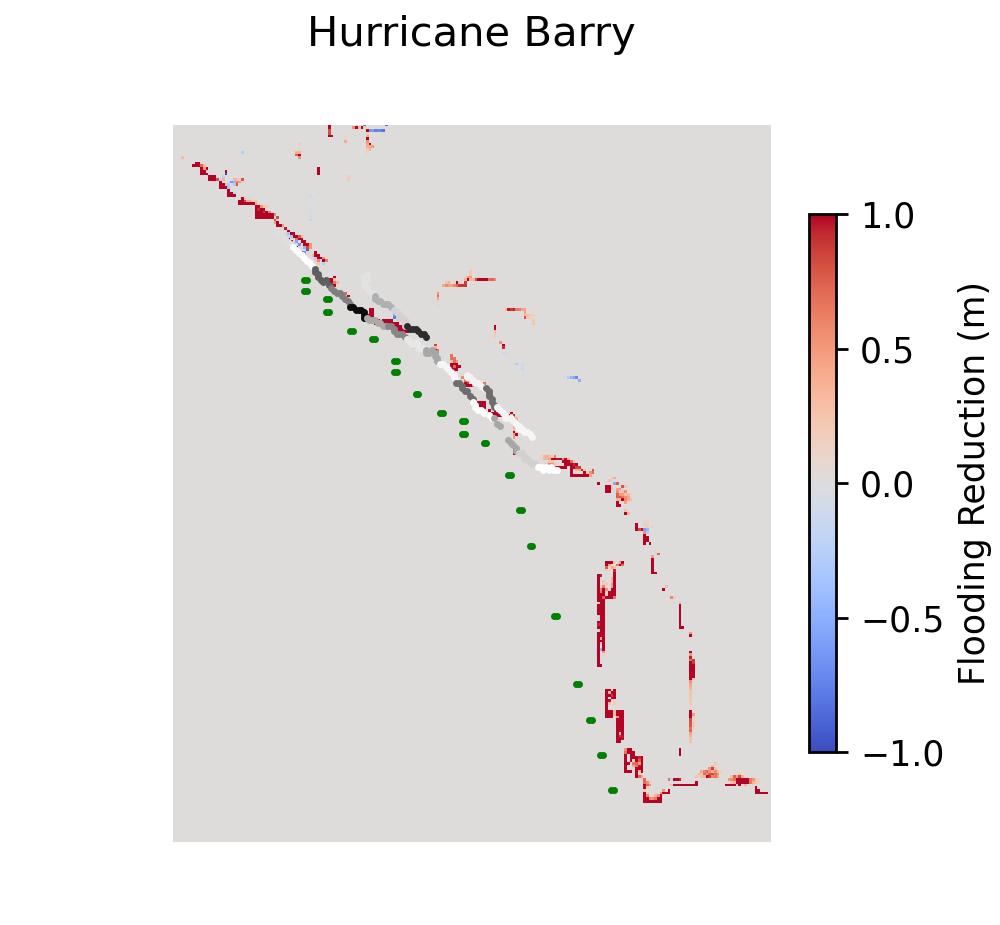}
\hfill
\includegraphics[width=0.325\linewidth]{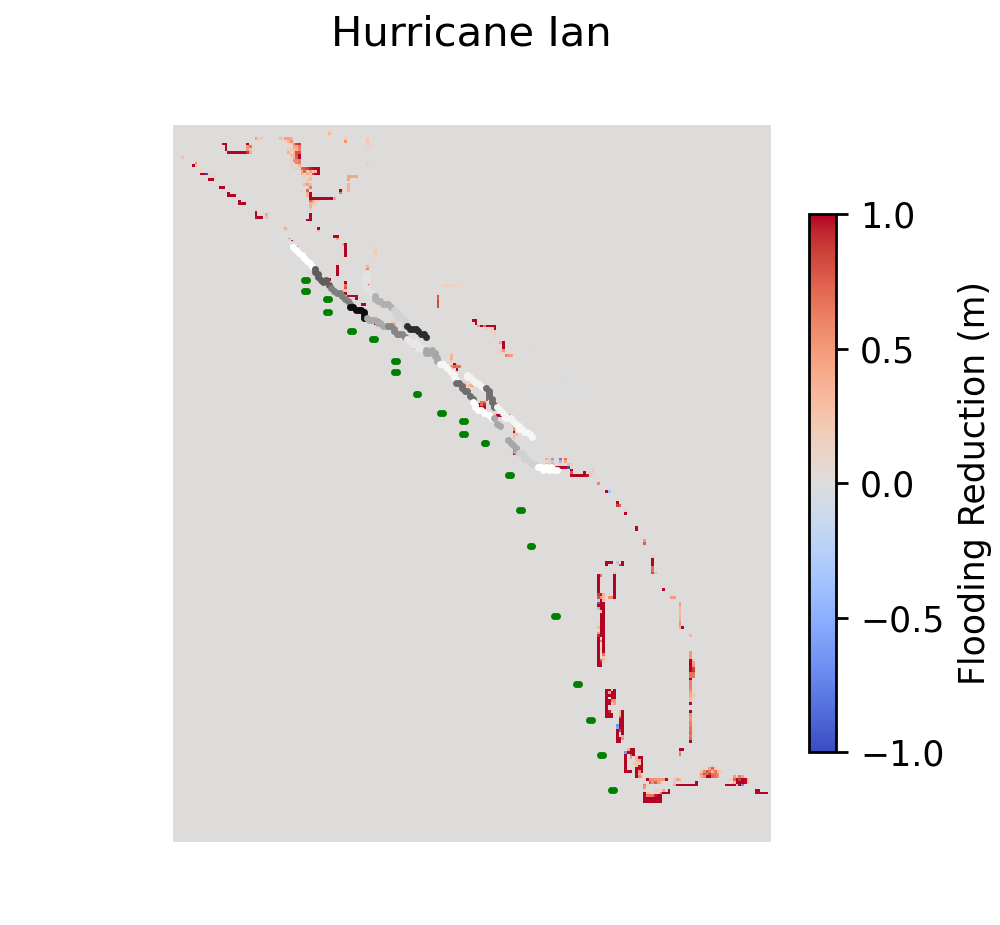}
\includegraphics[width=0.325\linewidth]{figures/bandit_results/idalia.png}
\hfill
\includegraphics[width=0.325\linewidth]{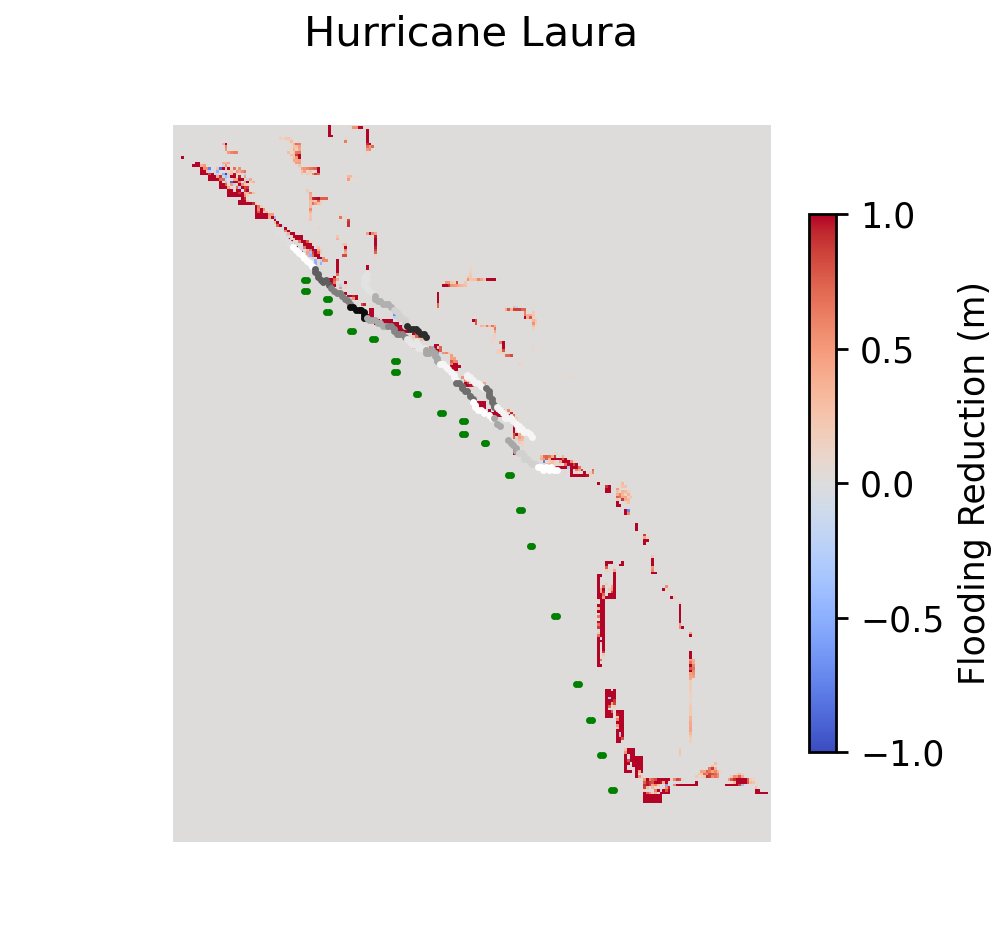}
\includegraphics[width=0.325\linewidth]{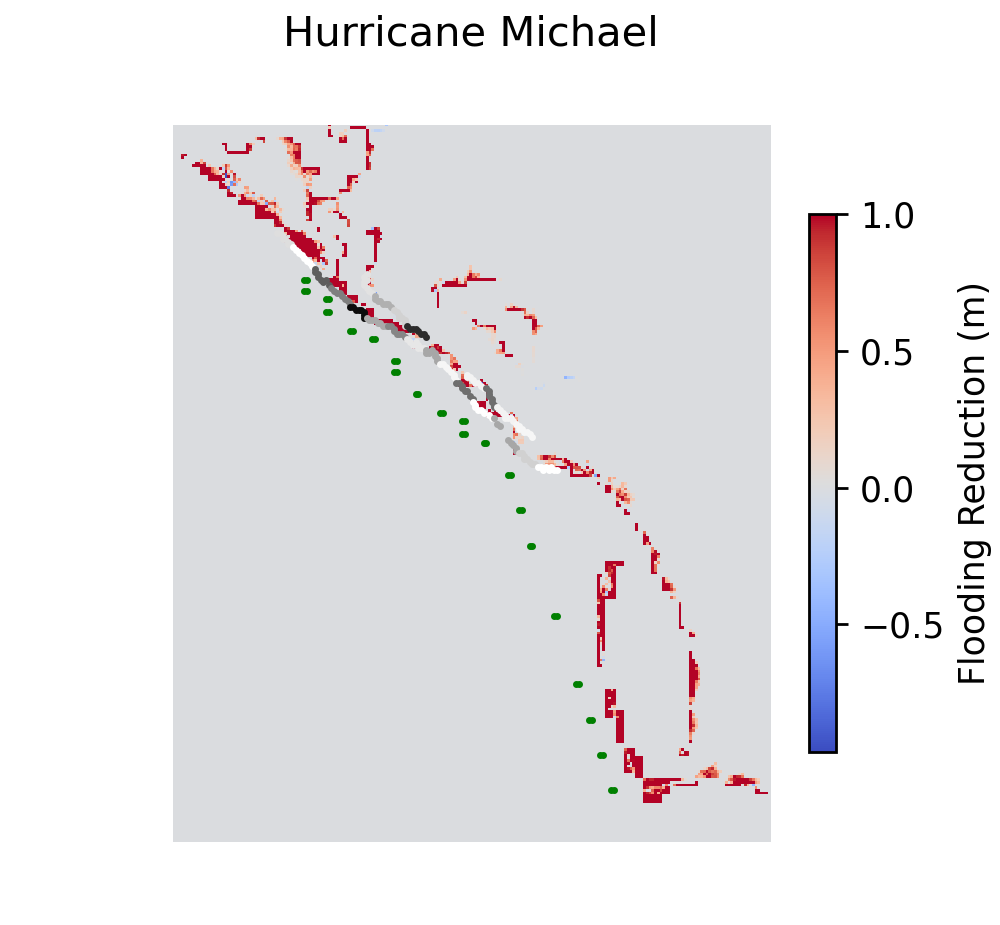}
\hfill
\includegraphics[width=0.325\linewidth]{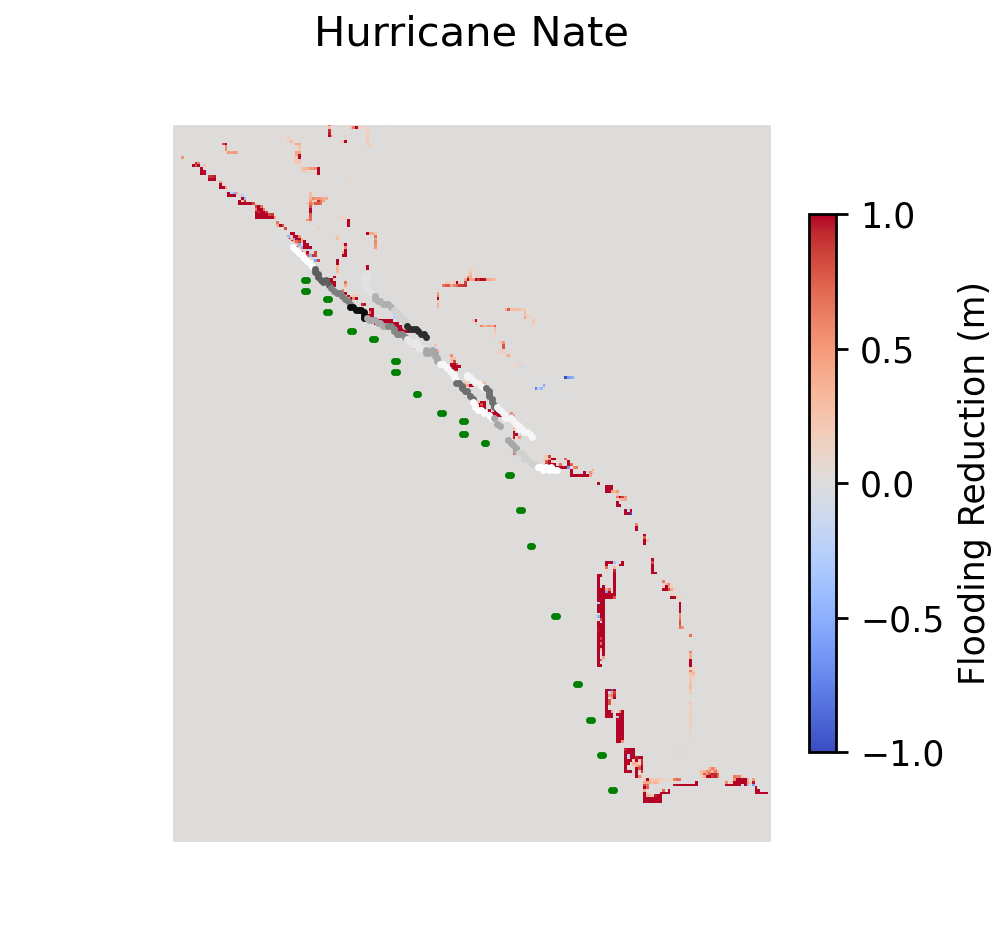}
\includegraphics[width=0.325\linewidth]{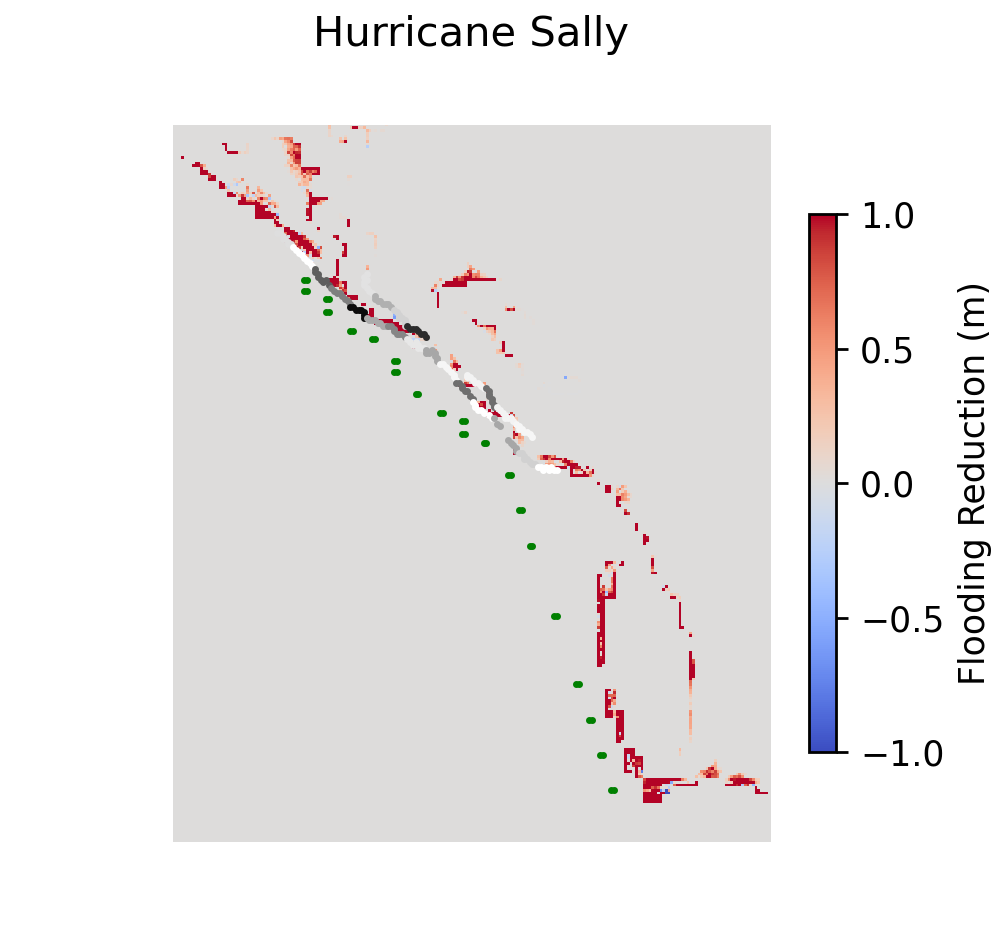}
\hfill
\includegraphics[width=0.325\linewidth]{figures/bandit_results/cost_mask.png}
\caption{
Optimized interventions. The green dots represent prescribed oyster reef sites and the white, grey, and black dots refer to the height of sea wall deployed at a potential site (dark is higher, up to 5m).  Red regions are where the interventions reduce flooding relative to no interventions; blue are where flooding is increased. Bottom right: cost mask of region, with colors reflecting costs on a logarithmic scale.}
\label{fig:placement_map_app}
\end{figure}

Here (\cref{fig:placement_map_app}) we include images of optimized intervention choices using the full ensemble of storms in the optimization.  We observe that resources are concentrated near the most populated area of Tyndall AFB, but are chosen to be generally effective at reducing flooding everywhere.  Note that the flooding reduction on all plots is clipped for visualization purposes.

%%%%%%%%%%%%%%%%%%%%%%%%%%%%%%%%%%%%%%%%%%%%%%%%%%%%%%%%%%%%

\end{document}